\documentclass{aa}  

\usepackage{graphicx}
\usepackage{natbib}
\bibpunct{(}{)}{;}{a}{}{,}

\newcommand{\micron}{$\mu$m}

\newcommand{\msun}{M$_{\sun}$}

\begin{document} 

   \title{The Massive Stellar Population of W49:\\
   A Spectroscopic Survey\thanks{Based on data acquired using the Large Binocular Telescope (LBT). The LBT is an international collaboration among institutions in Germany, Italy and the United States. LBT Corporation partners are: LBT Beteiligungsgesellschaft, Germany, representing the Max Planck Society, the Astrophysical Institute Potsdam, and Heidelberg University; Istituto Nazionale di Astrofisica, Italy; The University of Arizona on behalf of the Arizona university system; The Ohio State University, and The Research Corporation, on behalf of The University of Notre Dame, University of Minnesota and University of Virginia.},\thanks{Based on observations made with ESO Telescopes at the La Silla Paranal Observatory under programme IDs 67.C-0514 and 073.D-0837.}}

   \subtitle{}
   \titlerunning{The Massive Stellar Population of W49}
   \authorrunning{Wu et al.}

   \author{Shi-Wei Wu\inst{1}\fnmsep\thanks{International Max Planck Research School for Astronomy and Cosmic Physics at the University of Heidelberg (IMPRS-HD)},
          Arjan Bik\inst{2}\fnmsep\inst{1},
          Joachim M. Bestenlehner\inst{1},
          Thomas Henning\inst{1},
          Anna Pasquali\inst{3},
          Wolfgang Brandner\inst{1}
          \and
	Andrea Stolte\inst{4}
          }

   \institute{Max-Planck-Institut f\"{u}r Astronomie, K\"{o}nigstuhl 17, 69117 Heidelberg, Germany\\
              \email{shiwei@mpia.de}
             \and
             Department of Astronomy, Stockholm University, AlbaNova University Centre, 106 91 Stockholm, Sweden
            \and
             Astronomisches Rechen-Institut, Zentrum f\"ur Astronomie der Universit\"at Heidelberg, M\"onchhofstr. 12 - 14, 69120 Heidelberg, Germany        
             \and
             Argelander Institut f\"ur Astronomie, Auf dem H\"ugel 71, 53121 Bonn, Germany
             }


 
  \abstract
   {Massive stars form on different scales ranging from large, dispersed OB associations to compact, dense starburst clusters.
 The complex structure of regions of massive star formation, and the involved short timescales provide a challenge for our understanding of their birth and early evolution.
As one of the most massive and luminous star-forming region in our Galaxy, W49 is the ideal place to study the formation of the most massive stars.}
   {By classifying the massive young stars deeply embedded into the molecular cloud of W49, we aim to investigate and trace the star formation history of this region.}
   {We analyse near-infrared $K$-band spectroscopic observations of W49 from LBT/LUCI combined with $JHK$ images obtained with NTT/SOFI and LBT/LUCI. Based on $JHK$-band photometry and K-band spectroscopy the massive stars are placed in a Hertzsprung Russell diagram. By comparison with evolutionary models, their age and hence the star formation history of W49 can be investigated.}
   {Fourteen O type stars as well as two young stellar objects (YSOs) are identified by our spectroscopic survey. Eleven O-stars are main sequence stars with subtypes ranging from O3 to O9.5,  with masses ranging from $\sim 20$ \msun\ to $\sim 120$ \msun. Three of the O-stars show strong wind features, and are considered to be Of-type supergiants with masses beyond 100 \msun. The two YSOs show CO emission, indicative for the presence of circumstellar disks in the central region of the massive cluster. The age of the cluster is estimated as $\sim1.5$ Myr, with star formation still ongoing in different parts of the region. The ionising photons from the central massive stars have not yet cleared the molecular cocoon surrounding the cluster. W49 is comparable to extragalactic star-forming regions and provides us with an unique possibility to study a starburst in detail.}
   {}


   \maketitle
%

\section{Introduction}
Massive stars form in dense regions of giant molecular clouds (GMCs), and interact strongly with their environment. The environments where massive stars form range from dense starburst clusters to loose OB associations. The former are very compact regions with half-mass radii of one parsec or less and bound by self-gravity (e.g., \citet{Rochau}), while in the latter OB stars spread over scales from a few to tens of parsec. Such a difference in morphology and physical scale could have a strong influence on the early evolution of the star-forming regions and the stars within them. 
The near-infrared spectral window provides the possibility to detect radiation from the stellar photospheres of young massive stars, in spite of a visual extinction as high as $A_V \sim 50$ mag.

By investigating the stellar content of star-forming regions, we can try to understand how environmental effects, such as cloud morphology, or feedback by massive stars, influence the star formation history. We can address the question of whether clusters form in a single burst with stars all of the same age \citep{Kudryavtseva:2012aa}, or form over a longer time with star formation happening in different parts of a giant molecular cloud \citep{Blaauw:1991aa,de-Zeeuw:1999aa,Comeron:2012aa}. There is evidence that the Galactic Centre and the Galactic disk are assembling gas into massive clusters in different ways: in the disk, spiral arm density waves and large scale gas flows feeding a progenitor cloud via filaments seems to be the main mechanism. In the Galactic Centre, gas is able to reach very high density without forming stars until possibly cloud-cloud collisions or tidal forces trigger the collapse of gas under its own gravity \citep{Longmore:2014aa,Johnston:2014aa}. 

Massive stars are the main sources of ionising flux and mechanical energy (by means of stellar winds and supernova shock waves) injected into GMCs. By identifying and characterising the massive stellar content, the feedback on the surroundings can be studied in detail. An important question is if and under what circumstances the feedback by massive stars might trigger or quench further star formation \citep{Zinnecker:2007aa}. 

A study of the spatial distribution of massive stars also provides clues on the formation mechanisms of clusters and massive stars. It is still under debate if all massive stars form in clusters or if individual OB stars could form in isolation \citep{Bressert:2012aa,de-Wit:2005aa,Bonnell:2004aa,Banerjee:2012aa}.

Young massive clusters, where the majority of the very massive stars form and reside during their short life time \citep{Crowther2010aa,Bestenlehner:2011aa,Wu:2014aa}, are the best environment to study the physical conditions of the birthplaces and the early evolution of the most massive stars.

The LOBSTAR (Luci OBservation of STARburst regions) project is a near-infrared spectroscopic survey of the stellar content of several of the most massive star formation complexes in our Galaxy, including W3 Main, W49 and W51. In W3 Main \citet{Bik:2012aa} classified 15 OB stars and three YSOs, which is indicative of an age spread of at least 2 to 3 Myr between different subregions. The evolutionary sequence observed in the low-mass stellar population via photometry shows that W3 Main is still actively forming stars \citep{Bik:2014aa}. Nine OB stars and one YSO, associated with different \ion{H}{ii} regions in W51, have been identified by our spectroscopic classification (Wu et al., in prep). The wide spread, multiple episodes star formation has been found triggered and affected mainly by external effects such as galactic density wave and the current interaction with a supernova shock wave.

With dozens of massive stars in its core, W49 is one of the most important Galactic sites for studying the formation and evolution of the very massive stars. Given its location in the plane of the Milky Way and distance of {11.1 $^{+0.8}_{-0.7}$} kpc \citep{Zhang:2013ab}, W49 is optically obscured  by intervening interstellar dust, and subject to large amounts of crowding and field star contamination by foreground stars. 

Using deep near-infrared imaging, \citet{Alves:2003aa} and \citet{Homeier:2005aa} studied the stellar population and the mass function, and reported the detection of massive stellar clusters still deeply embedded in the GMC of the W49 complex. The observations reveal high extinction towards W49, and large internal extinction variation.  At least $A_V > 20 $ mag of foreground extinction and more than $30$ mag of internal inhomogeneous extinction were found in this region. They derive a total stellar mass of $5-7 \times 10^4 $ \msun, which makes W49 comparable to extragalactic giant star-forming regions.

W49 was also the subject of several radio and submillimeter studies \citep{de-Pree:1997aa,Roberts:2011aa,Nagy12,Galvan-Madrid:2013aa,Nagy15}, which revealed complex kinematics of the molecular gas in W49, with a mixture of infall and outflow motions. There are several clumps of cool and dense gas surrounding, and possibly infalling onto the centre of the region \citep{Roberts:2011aa}. With only $1\%$ of the gas being photoionized, star formation in W49 is ongoing and the feedback from the cluster is not (yet) strong enough to halt the process \citep{Galvan-Madrid:2013aa}. As comparable physical conditions have been measured in extragalactic starburst regions, W49 could serve as a template for the luminous, embedded star clusters being found in normal and starburst galaxies. 

In the following sections, we present our near-infrared observation of W49 from LBT/LUCI and SOFI/NTT. The reduction of the imaging and spectroscopic data is presented in Sect.\ 2; in Sect.\ 3 we derive the astrophysical properties of the massive stars, and place them in a Herzsprung Russel diagram (HRD); the fundamental properties of the cluster in W49, its formation history, feedback towards the environment and the spatial distribution of massive stars are discussed in Sect.\ 4. The result of our spectroscopic investigation towards W49 is summarised in Sect.\ 5.

\section{Near-infrared observations and data reduction}

\begin{figure*}
   \centering
   \includegraphics[width=0.9\hsize]{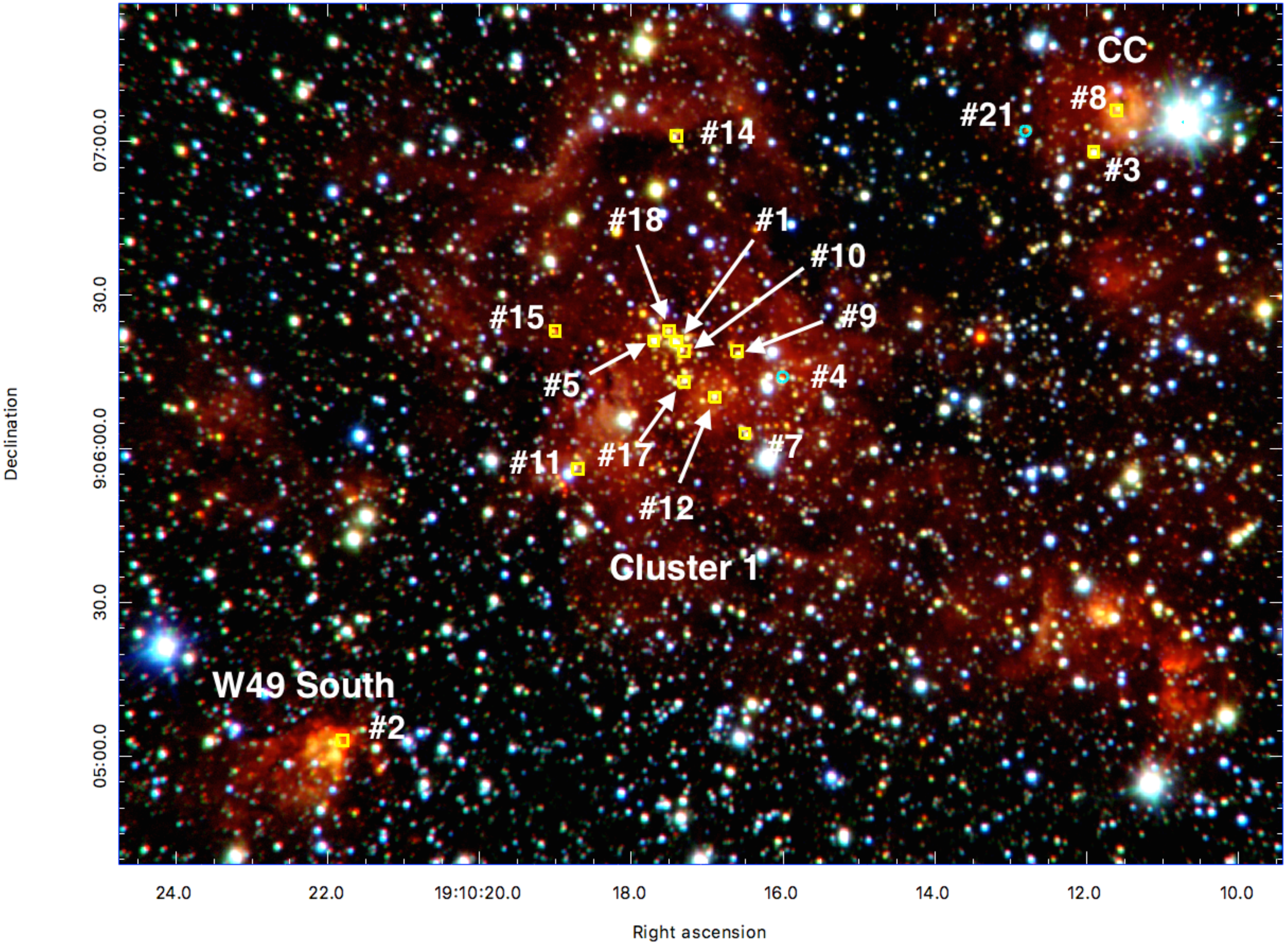}
      \caption{ $JHK$ composite color image of the central 12\,pc $\times$ 9\,pc of W49. North is up, and east is to the left. The red, green, and blue channels are mapped to the $K$, $H$, and $J$, respectively. The $J$ and $H$-band image are from SOFI/NTT and the $K$-band image is from LUCI/LBT. The spectroscopically identified OB-type stars are marked by yellow squares. Two Young Stellar Objects are marked with cyan circles. Two radio sources (W49 South and CC, also associated with sub-clusters) are labeled in white.}
         \label{W49JHK}
   \end{figure*}

The observations have been carried out with LUCI mounted on the Large Binocular Telescope \citep[LBT, ][]{Hill06}, Mount Graham, Arizona. LUCI is a near-infrared multi-mode instrument capable of Multi-Object Spectroscopy (MOS), long-slit spectroscopy and imaging \citep{Seifert10,Ageorges10,Buschkamp10}. The spectra of the massive stars in W49 have been taken in MOS mode based on   K-band pre-image also obtained with LUCI. Additional archival data were used to complement the LUCI data.  Medium-resolution (R=10,000) $K$-band spectra of five massive stars in W49 obtained with ISAAC mounted on Antu (UT1) of ESO's Very Large Telescope (VLT), Paranal, Chile, and  $J$- and $H$-band images obtained with SOFI at the New Technology Telescope (NTT), La Silla, Chile, were downloaded from the ESO archive.
  
\subsection{Observations}

\subsubsection{Imaging Observations}

The K-band image of W49 was taken on 2009, September 29 with the N3.75 camera of LUCI/LBT with a total exposure time of 840 s. More details on the imaging observations can be found in \citet{Wu:2014aa}.

The archival $J$ and $H$-band data \citep[first published in ][]{Alves:2003aa} were downloaded from the ESO archive and are the same as used in \citet{Wu:2014aa}. The observations were  performed on 2001, June 7 with SOFI/NTT with a total exposure time of 600\,s and 450\,s in $J$ and $H$, respectively. All data were taken under good atmospheric conditions with a typical angular resolution of 0.5\arcsec\ to 0.7\arcsec.  
The effective area covered by all three bands is 5\arcmin $\times$ 5\arcmin.

\subsubsection{Source selection for  spectroscopic observations}

 Several selection criteria were applied to select the targets for the followup spectroscopy with LUCI. First, following \citet{Alves:2003aa}, we selected sources with  $H-K > 1.2$\,mag as potential cluster members. Only red-wards of this color \citet{Alves:2003aa} are able to detect the embedded clusters. In order to minimize the fore- and background contamination even more, we selected only sources which are associated with one of the 4 clusters in W49.  The large list of candidate cluster member stars was then used to create the masks for the MOS observations.

Not all stars could be observed using the MOS masks as slits on other stars prevented their selection. Our completeness is dominated by the design of our MOS masks and is worse in the center of the clusters than in the outskirts. Also we added additional stars to fill in empty places on the masks in regions where no candidate cluster members were left. These additional stars typically violated the selection criteria as described above.

\subsubsection{Spectroscopic Observations}

We observed W49 with the MOS mode of LUCI in the $K$-band from 2010 May 14 to June 11 and from 2011 April 11 to May 15 under varying atmospheric conditions. We used the 210\_zJHK grating and slit width of 0.7\arcsec\ for the masks targeting the brighter stars, and 1\arcsec\ for the remaining stars.  The angular sampling of the spectra is $0.''25~pixel^{-1}$ with the N1.8 camera, which provides the largest wavelength coverage ($\Delta\lambda=0.328\mu$m). 
 
In addition to the MOS spectra from LUCI, we reduced a set of archival long-slit spectra taken with ISAAC in the $K$-band on 2004 August 6, providing a  wavelength coverage between 2.08 $\mu$m and 2.20 $\mu$m. The spectra were collected with a angular sampling of $0''.147~pixel^{-1}$. The $K$-band spectrum of the very massive star W49nr1 \citep{Wu:2014aa}, is also part of this dataset. More details on the  ISAAC spectra are presented in \citet{Wu:2014aa}.

\subsection{Data reduction}
The reduction of the imaging data and the ISAAC long-slit spectra is described in \citet{Wu:2014aa}. 
The LUCI spectra were reduced with a modified version of \emph{lucired}, which is a collection of IRAF routines developed for the reduction of LUCI MOS spectra. The raw frames were first corrected for the tilt of the slit and distortion using  spectroscopic sieve and imaging pinhole masks, respectively. The science and standard star spectra were divided by the normalized flat field. The MOS spectra were cut into individual slits and the wavelength calibration was carried out using the Ar and Ne wavelength calibration frames. After the wavelength calibration, the sky background was removed by subtracting two frames adjacent in time or using the procedure by \citet{Davies:2007aa} (which corrects for the variations of OH lines by fitting the individual transitions to minimize the residuals), depending on which one of these two methods was more successful in minimizing sky-lines residuals. Then the one-dimensional spectra were extracted using the IRAF task \emph{doslit}. The local background was estimated by fitting a region close to the star with a Legendre function, so that the narrow $Br\gamma$ emission from the surrounding diffuse nebular structure can be removed from the final spectra. At last the individual exposures for each star are combined into the final spectra.

In preparation for telluric correction,  the $Br\gamma$ absorption line in the spectrum of the telluric standard star was removed by fitting the line with a Lorentzian profile. The resulting atmospheric transmission spectrum was used with the IRAF task \emph{telluric} to correct the science spectra. After comparing the science spectra corrected with the telluric standard stars taken before and after the science exposure, we selected the science spectra with the least telluric residuals. Finally, the spectra of the science targets are normalized and presented. 
A residual $Br\gamma$ emission component remains due to intensity variations of the nebular emission on small spatial scales.

In total, good quality spectra of 44 stars have been obtained with identifiable features in their continuum. Half of them have spectra dominated by CO absorption bands and other atomic absorption lines (Table.~\ref{Alltable}). They are identified as late-type foreground dwarf and giant stars and are not members of W49 and therefore not discussed in this paper. The remaining 22 stars (Table.~\ref{table:1}) show the spectral features of massive stars and  YSOs and are candidate members of W49. In the following, their membership to W49 is discussed in more detail.

\begin{figure}
   \includegraphics[width=\hsize]{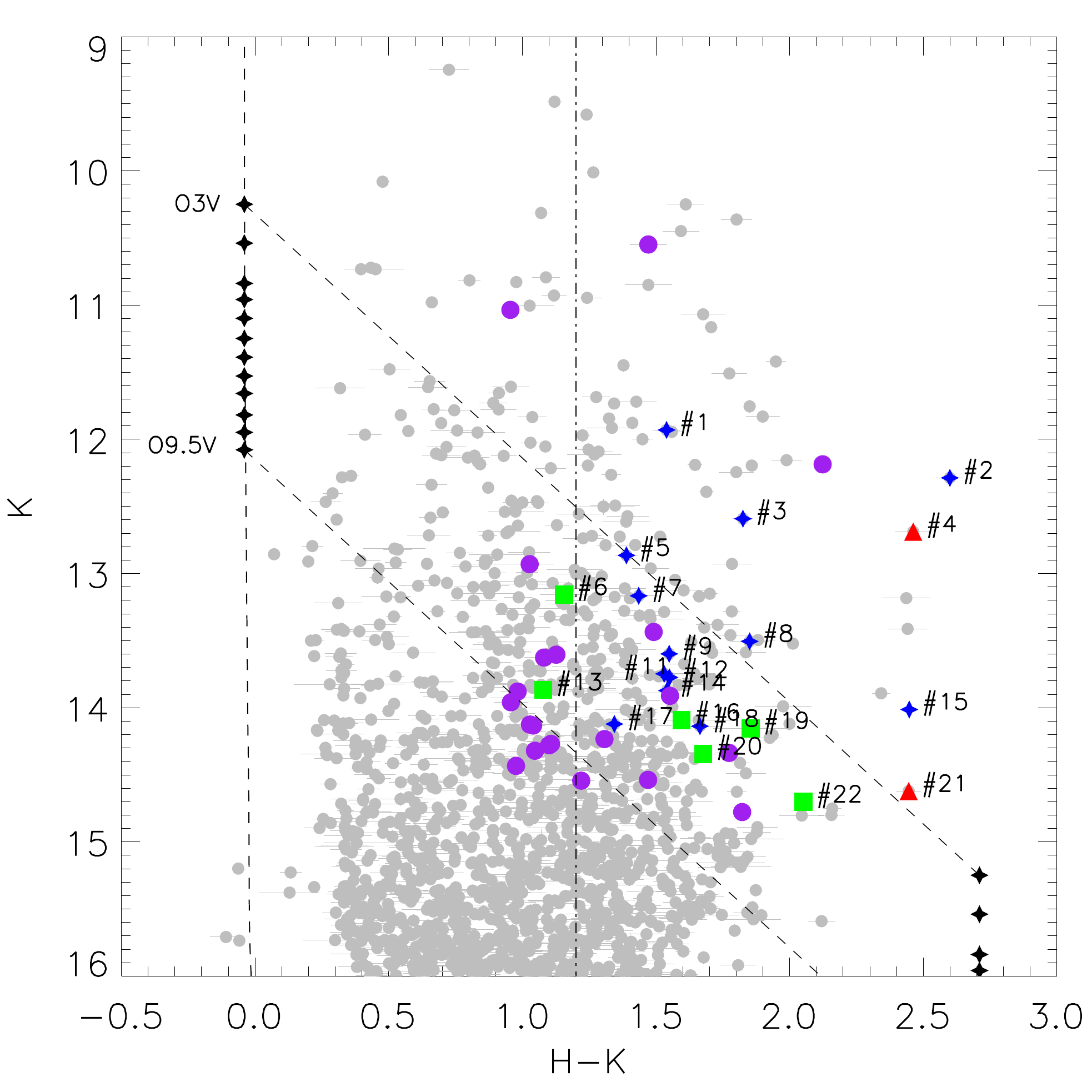}
      \caption{$H-K$ vs. $K$ color-magnitude diagram of W49. All sources identified in $JHK$, with errors less than 0.1 mag and with $K \le 16$\,mag are plotted in grey. The blue diamond symbols represent the spectroscopically identified OB stars,  red triangles are the YSOs with CO emission in their spectra, while the green squares mark the location of the stars with $Br\gamma$ absorption lines as the most obvious feature,  which cannot be classified due to a lack of other features (see discussion in the text). The purple circles are the 22 late type stars idenfied as fore- or background stars. The dashed vertical line represents the un-reddened isochrone for main sequence stars with an age of 1\,Myr \citep{Ekstrom:2012aa,Yusof:2013aa}. The black star symbols on the dashed vertical line mark the O stars with subtypes O3V to O5V with a step of 1, and O5V to O9.5V with a step of 0.5. The black star symbols in the bottom right mark the locations of O stars of different subtypes for $A_{K_{\rm{s}}}$ =5\,mag. The two diagonal dashed lines mark the reddening directions of an O3V and O9.5V star according to the extinction law of \citet{Indebetouw:2005aa}.
    \label{CMD}}
\end{figure}

\begin{figure}
   \includegraphics[width=\hsize]{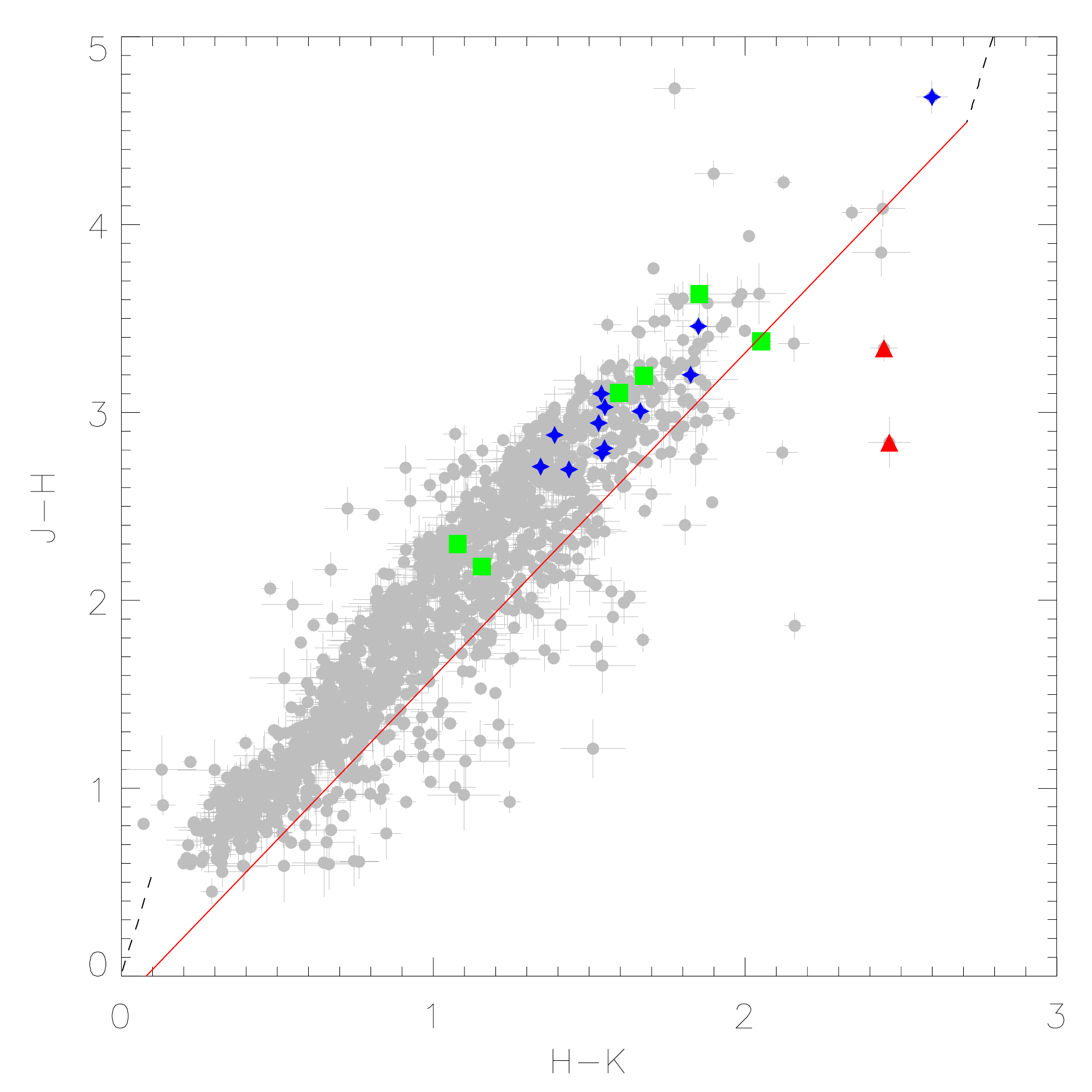}
\caption{$J-H$, $H-K$ color-color diagram of W49 for only the 22 sources discussed in the paper. The late type stars are not included here. The symbols have the same meaning as described in Fig~\ref{CMD}. The two black dashed lines represent main sequence isochrones with the age of 1 Myr and the initial mass ranges from 0.8 to 500 \msun\  from \citet{Ekstrom:2012aa} and \citet{Yusof:2013aa} without being reddened (bottom left) and being reddened with $A_{K_{\rm{s}}}$ =5\,mag (upper right). The red diagonal line represents the reddening law according to \citet{Indebetouw:2005aa}. 
    \label{CCD}}
\end{figure}

\section{Results}

In this section we present the near-infrared photometry as well as the $K$-band spectroscopy data for W49. Based on both the imaging and spectroscopy we identify different classes of objects and derive the spectral type of the identified massive stars.

\subsection{Photometric classification}

Near-infrared imaging of the W49 star-forming region shows that this region is dominated by a dense central cluster surrounded by smaller sub-clusters (Fig~\ref{W49JHK}).  Due to the large distance of W49, high extinction and high foreground contamination make it impossible to reliably determine cluster membership on photometry alone. This is clearly demonstrated by the observed $H-K, K$ color-magnitude diagram  (CMD, Fig~\ref{CMD}), showing a continuous spread in color, without a clearly identifiable reddened main sequence as the reddened cluster population. The  $J-H$, $H-K$ color-color diagram (CCD, Fig~\ref{CCD}) shows that the large range in  $H-K$ color corresponds to an extinction range $A_{K}$ = 0 to 5\,mag when applying the extinction law of \citet{Indebetouw:2005aa}. 

To get an idea of the stellar population in W49, \citet{Alves:2003aa} and \citet{Homeier:2005aa} applied a $H-K \ge 1.2$\,mag color cut (equivalent to $A_{K} $ = 2.1\,mag), based on the clustering of the stars.  As described in Sect. 2,  we selected stars in a similar fashion for our spectroscopic observations. In the remainder of this section we focus on the photometric properties of the stars selected for spectroscopy.

{The observed $JHK$ magnitudes of the spectroscopically observed stars are listed in Table \ref{table:1}.}  Almost all stars are detected in $JHK$, allowing a characterization of the sources using the CMD and CCD. Exceptions including source \#10, which is only detected in $K$, and is blended with brighter stars in $J$ and $H$. The crowding in the $K$ band still results in a high photometric uncertainty. Source \#15 is not detected in $J$ as this source is highly reddened and hence too faint. 

All the spectroscopically observed stars are marked in the CMD and CCD, as well as the applied $H-K \ge 1.2$\,mag color cut. The locations of the spectroscopically identified massive stars (see section 3.3) show that indeed the cluster is extremely reddened and that extinction within W49 is highly variable.

In the CMD, the black star symbols mark the positions of different subtypes of main sequence stars ranging from O3V to O9.5V taken from \citet{Martins:2006aa},  adopting the extinction law of \citet{Indebetouw:2005aa} and assuming a distance of 11.1 kpc \citep{Zhang:2013ab} for W49. The reference points are then reddened by $A_{K_{\rm{s}}}=5$\,mag. The spectral types of the candidate massive stars can be estimated by comparing the positions of the observed stars in the CMD with the reddened main sequence. The resulting photometric spectral types are given in Table~\ref{SpectralType} and can be compared to our spectral classification based on the LUCI spectra (Section 3.3). Most stars are classified between O3V and O9.5V, suggesting that they are massive stars inside W49. Some stars appear to be more luminous than a single O3V star. In these cases a spectral classification is mandatory to reveal the true nature.

As discussed in \citet{Wu:2014aa} the choice of the extinction law will have a significant effect on the de-reddened magnitudes and therefore on the photometric spectral type. We use the extinction laws of \citet{Indebetouw:2005aa,Fitzpatrick99,Nishiyama09}  and \citet{Rieke85}, whose slopes are consistent with the observed colors in W49, to estimate the uncertainty due to different extinction laws on the photometric spectral type. Different extinction laws result in 2 to 3 subtype uncertainty in the photometric spectral type determination. The extinction law by \citet{Fitzpatrick99} yields earlier subtypes, \citet{Nishiyama09} yields later subtypes, while \citet{Indebetouw:2005aa} and \citet{Rieke85} give comparable results in between the previous two.


\begin{table}
\caption{Spectral types of the massive stars in W49: Photometric classification derived from the  CMD  and spectroscopic classification from the LUCI $K$-band spectra compared to the mid-infrared photometric classification of \citet{Saral:2015aa}; the numbering of the massive stars is according to their $K$-band magnitude from bright to faint.}             
\centering          
\begin{tabular}{l r r l}
\hline\hline       
Star &  Ph. Class. & Sp. Class. & \citet{Saral:2015aa} Class. \\ 
\hline                    
\#1  & $<$O3 & O2-3.5If* & class III/photosphere \\
\#2  & $<$O3 & O2-3.5If* & class I\\ 
\#3  & $<$O3 & O3-O7V & class III/photosphere\\ 
\#4  & O$^\dag$& YSO  &class I\\
\#5  & O3-O4 & O3-O5V &unclassified\\ 
\#6  & O5.5-O6& B2-B3V &unclassified\\ 
\#7 & O3-O4 & O3-O5V  &unclassified\\
\#8  & $<$O3 & O3-O7V  &class I\\
\#9  & O4-O5 & O3-O7V &--- \\
\#10  & - & O5-O7V & ---\\ 
\#11  & O5-O5.5 & O3-O5V & ---\\ 
\#12  & O5-O5.5 & O5-O7V &---\\ 
\#13  & O8-O8.5 & B0-B2V  &---\\
\#14  & O5.5-O6 & O3-O5V &class I\\ 
\#15  & $<$O3 & O2-3.5If* & class II\\ 
\#16  & O6-O6.5 & $Br\gamma$ abs &---\\
\#17  & O7.5-O8 & O7-O9.5 &---\\ 
\#18  & O5.5-O6 & O8-O9.5 &---\\
\#19  & O4-O5& $Br\gamma$ abs &---\\ 
\#20  & O6-O6.5& $Br\gamma$ abs&---\\
\#21  & O$^\dag$ & YSO &unclassified\\
\#22  & O5-O5.5& $Br\gamma$ abs &--- \\ 
\hline
\label{SpectralType}
\end{tabular}
\tablefoot{
\tablefoottext{\dag}{Luminous YSOs with intrinsic IR excess.}
}
\end{table}

\subsection{Excess sources}

\begin{figure*}
   \centering
   \includegraphics[width=0.9\hsize]{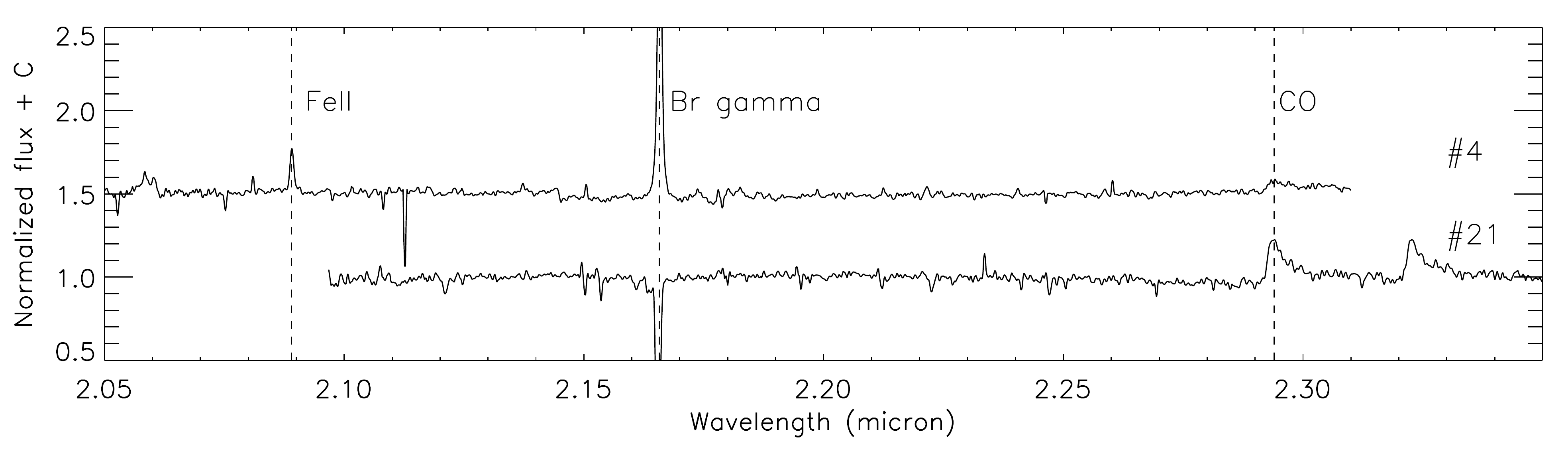}
   \caption{Normalized $K$-band spectrum of two stars identified as YSOs in W49. The most important features are the CO emission lines at the long wavelength end of their $K$-band spectra.} 
   \label{YSOSpectra}
\end{figure*}

In addition to stars located on or near the reddening line of the CCD (Fig~\ref{CCD}), several sources are located to the right of the reddened main sequence. These stars possibly possess an intrinsic infrared excess due to circumstellar material.
In general the fraction of infrared excess sources is a strong function of the age of a stellar cluster  \citep[e.g.][]{Hernandez08}. Additionally, dispersion of circumstellar material is driven by external factors like photo evaporation \citep{Hollenbach:2000aa} or dynamical interactions with surrounding stars \citep{Olczak:2010aa}, resulting in a much lower disk fraction in massive stellar clusters \citep[e.g.][]{Stolte10,Fang12,Bik:2014aa,Stolte15}. 

For W49 we cannot determine a reliable disk fraction, as with the current data we cannot separate the cluster members from the fore- and background stars. However we can identify individual YSOs and discuss their likelihood to be a member of the cluster. Two of the most extreme excess sources in the CCD are covered by our LUCI spectra (Fig~\ref{YSOSpectra}).  Both sources (\#4 and \#21) have an extremely red color $H-K$ > 2.4\,mag. 
Source \#4 has been identified by \citet[their nr 2]{Conti02} as a candidate OB star, by \citet{Homeier:2005aa} as a candidate for very massive star with a mass in excess of 120\,M$_\odot$,  and by \citet[nr 7 in their Table 10]{Saral:2015aa} as a candidate YSO with a mass estimate of $\approx 8.9$\,M$_\odot$ (Tab. \ref{SpectralType}). We note that the latter result is based on SED fitting and relies on 2MASS and Spitzer/IRAC photometry, and hence might be affected by crowding and blending.
The YSOs are marked by red triangles in the CMD and CCD of Fig~\ref{CMD} and Fig~\ref{CCD}. Their location in the upper right corner of the CCD suggests that they have a similar extinction as the candidate O stars in W49 and that their extreme  $H-K$ color is caused by the infrared excess emission.

The $K$-band spectra show the CO $\nu$ = 2-0 and 3-1 overtone bands at around 2.3 \micron\ in emission, which is frequently observed in YSOs \citep{Bik:2004aa,Bik:2006aa,Stolte10,Wheelwright10}. In addition to CO, star \#4 also shows $Br\gamma$ and \ion{Fe}{II} 2.089 \micron\ emission. The $Br\gamma$ absorption line in the spectrum of \#21 is caused by over subtraction of the diffuse nebular $Br\gamma$ emission.

The CO emission likely arises from the neutral($\sim$ 2000\,K) dense inner regions of the disk \citep{Bik:2004aa,Wheelwright10}, while the $Br\gamma$ emission is originating from the surrounding \ion{H}{II} region as well as the ionised regions of the circumstellar disk. The \ion{Fe}{II} emission in the spectrum of \#4 is seen in several high-luminosity objects and could be caused by  UV fluorescence \citep{Mcgregor88}.  

As we do not know the relative contribution of the disk and the star to the total observed flux we cannot determine the spectral type of the underlying star via photometry. The presence of $Br\gamma$ and \ion{Fe}{II} lines in the spectrum suggests that the central source is a hot and probably massive star. Similar to \citet{Bik:2006aa}, we try to estimate the spectral type of the underlying star by comparing its location in the CMD with that of well studied massive YSOs. \citet{Bik:2006aa} constructed a CMD from the de-reddened $J$-$K$ color and the absolute $K$ magnitude (their Fig~1). By applying an average extinction towards W49 of $A_{K}$ = 3 mag (Table \ref{OBtable}) and a distance modulus of 15.22 mag, we derive absolute magnitudes of M$_{K}$ = -5.5\,mag and M$_{K}$ = -3.9\,mag for \#4 and \#21 respectively. The de-reddened $J$-$K$ colors are 0.8 mag (\#4) and 1.3 mag (\#21). Placing the objects into the diagram of \citet{Bik:2006aa} shows that their central stars are most likely late O stars and that they have a rather blue SED with little dust present. \citet{Bik:2006aa} explain this as a result of disk dispersal. The outer regions of the disk are dispersed faster than the inner region and a small and hot  inner disk remains before the disk is totally destroyed by the UV photons.


\subsection{Spectral classification of the massive stars}

Our total spectroscopic sample consists of 44 sources (Table.~\ref{table:1} and \ref{Alltable}). 22 of those sources have spectra dominated by CO absorption bands and other atomic lines. These stars are of spectral type G or later and are identified as foreground or background stars and not further analyzed.

\begin{figure*}
   \centering
   \includegraphics[width=0.9\hsize]{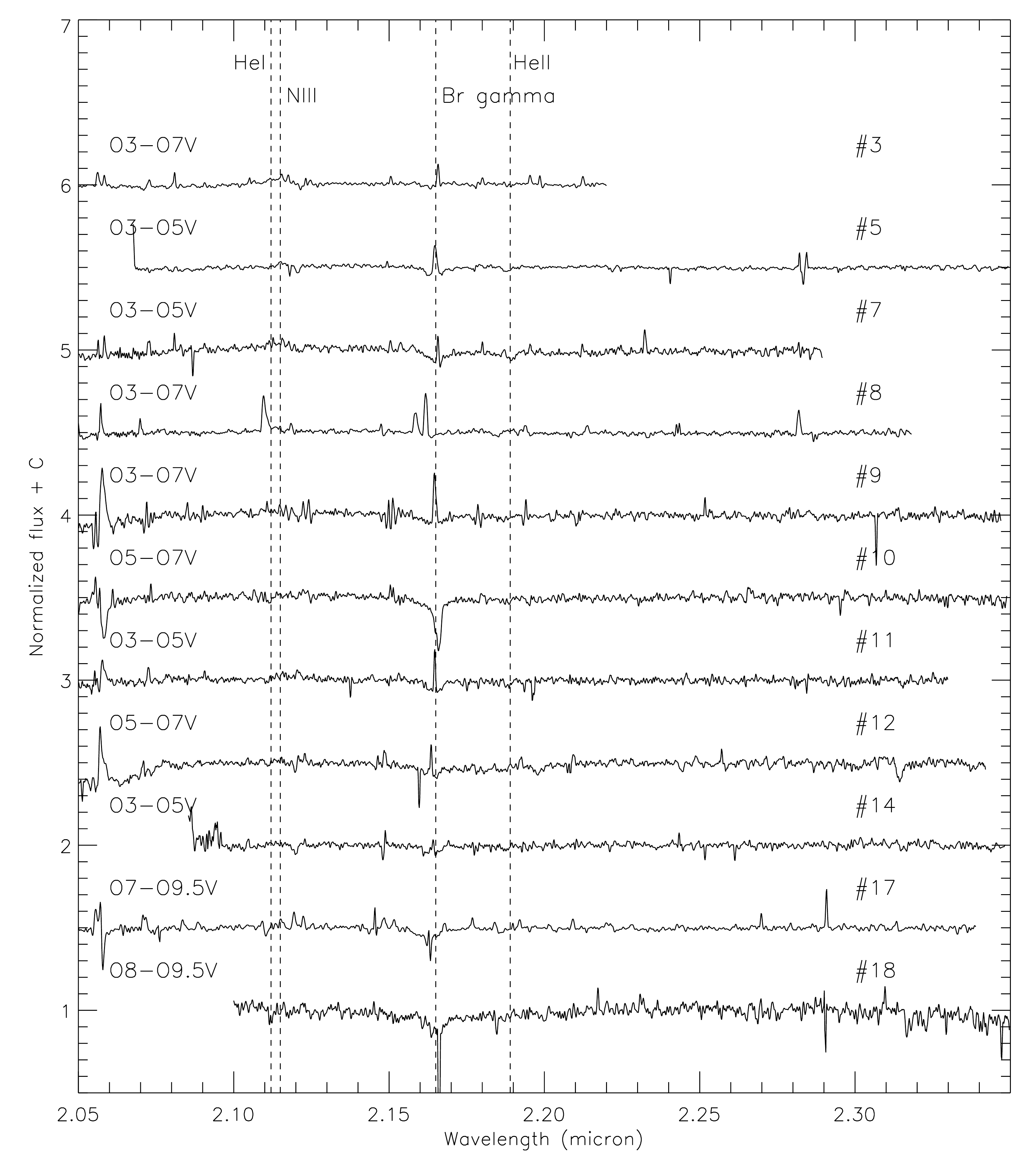}
   \caption{Normalized $K$-band spectrum of OB type stars in W49 as taken with the multi-object mode of LUCI. The star number and the spectral type based on the spectroscopic classification are indicated above the corresponding spectrum. Indicated with dashed lines are the spectral features crucial for spectroscopic classification.}
   \label{OBSpectra}
\end{figure*}

Of the remaining 22 stars (Table.~\ref{table:1}), the two YSOs have been discussed in the previous section. The other 20 stars show spectral features typical of  OB stars. 11 of them show absorption line spectra typical for O main sequence stars (Fig.\ \ref{OBSpectra}). Two stars show B type spectra and 4 stars have low SNR spectra and can not be classified properly. In section 3.3.1 we discuss the classification of these objects in more detail. Three remaining objects show emission line spectra suggestive of a stellar wind (Fig.\ \ref{OfSpectra}). Their classification is discussed in Section 3.3.2.

\subsubsection{OB main sequence stars}

The spectra of the OB stars all show $Br\gamma$ absorption originating in the stellar photosphere. Some $Br\gamma$ profiles show a narrow emission component in the center originating in the surrounding \ion{H}{II} region. As discussed in the data reduction section, we tried to correct the spectra for nebular emission by fitting and subtracting the background. Nevertheless, residuals are still left in some spectra because of the small-scale variations of the nebular emission. Other lines used for the derivation of the spectral type are the 2.10\micron\ (\ion{N}{V}), 2.113\micron\ (\ion{He}{I}), 2.116\micron\ (\ion{N}{III}) and 2.189\micron\ (\ion{He}{II}) lines.

The LUCI spectra of the massive stars have been visually compared to high resolution $K$-band spectra of reference O and B type stars with optical classification from \citet{Hanson:1996aa,Hanson:2005aa} and \citet{Bik:2005aa}. The high resolution spectra of \citet{Hanson:2005aa} and \citet{Bik:2005aa} are rebinned to the resolution of the LUCI spectra and artificial noise was added to the reference spectra to degrade them to the S/N level of the LUCI spectra. The errors on the spectral types are derived where visual comparison shows clear mis-matches between the observed and reference spectra.  Typically, the error is 1 or 2 sub types. 

As shown in Table~\ref{OBtable}, 11 out of the 22 stars are identified as OB main sequence star. Almost all spectra are classified as early type O stars (earlier than O7V) as they all show \ion{He}{II} and \ion{N}{III} in their spectra, indicative of a high effective temperature. To better refine the spectral classification we used the absence of the \ion{He}{I} (2.11\micron) line as an indicator for stars with spectral types between O3V and O5V. The strength of the $Br\gamma$ and HeII absorption lines varies from star to star \citep[see also][]{Hanson:1996aa,Hanson:2005aa}, probably due to stellar wind variations. Therefore matching the strength in the spectra with those of reference spectra will not result in a good spectral type estimate.  

Some stars (e.g. \#3 and \#8) show strong contamination by nebular emission of \ion{He}{I}, and therefore this line cannot be used as discriminant between early O and mid O stars. Two stars (\#6 and \#13) are classified as early B stars. 


Four relatively faint stars among our sample (\#16, \#19, \#20 and \#22) have low S/N. No obvious features except the broad $Br\gamma$ absorption lines are found in their spectra. This would classify them as early type, however a more detailed classification is not possible. As they are fainter than the stars classified above, we suspect that they are lower-mass cluster members and possibly late-O or early-B stars.

Five stars are observed both with LUCI and with VLT/ISAAC and are marked by \dag\ symbols after their name in Table~\ref{OBtable}. 
The first one is the very massive star reported in \citet[W49nr1]{Wu:2014aa}. The other four were identified as O3V to O5V stars according to their ISAAC spectra. The ISAAC spectra confirm the spectral classification derived from the LUCI spectra.

\begin{table*}
\caption{Physical parameters of the OB stars}             
\centering          
\begin{tabular}{l@{~~} l@{~~} l@{~} c@{~} c@{~} c@{~} c@{~} c@{~} c@{~} c}
\hline\hline       
Nr. &  $A_{k}$ & Sp. Type & Sp.Type & log$T_{eff}$  & log$L$ & Mass & Upper\tablefootmark{c} & Lower\tablefootmark{c}  &  log$Q_0$\\ 
	& (mag)	& (Hanson)\tablefootmark{a}  & (Crowther)\tablefootmark{b} & (K) & ($L_{\sun}$) & (M$_{\sun}$) & (M$_{\sun}$) & (M$_{\sun}$)& ($s^{-1}$) \\
\hline                    
\#1$^\dag$  &  $2.83\pm0.29$$^\ddag$ & O3If*-O4If  &   O2-3.5If*& $4.65\pm{0.02}$$^\ddag$ & $6.28\pm{0.15}$$^\ddag$ & $130\pm30$$^\ddag$ & 130 &50& 50.03-50.10$^\ddag$ \\
\#2  &    $4.78\pm0.53$$^\ddag$  & O4If-O5.5If  &  $>$ O3.5If* & $4.55\pm{0.02}$$^\ddag$ & $6.64\pm{0.25}$$^\ddag$  & $250\pm120$$^\ddag$ &240&90& 50.03-50.27$^\ddag$ \\ 
\#3  &  $3.50\pm0.04$ & O3-O7V &  - & $4.61\pm{0.05}$ & $6.18\pm{0.13}$ & $105\pm20$  &-&-& 49.58-50.11 \\ 
\#5$^\dag$  &  $2.71\pm0.13$ & O3-O5V & -  &  $4.63\pm{0.02}$ & $5.82\pm{0.10}$ &$64\pm8$   &-&-& 49.47-49.72 \\ 
\#7$^\dag$  &   $2.79\pm0.08$  & O3-O5V & -  & $4.63\pm{0.02}$ & $5.73\pm{0.09}$ & $57\pm5$   &-&-& 49.39-49.62 \\ 
\#8$^\dag$  & $3.55\pm0.17$ & O3-O7V & -  & $4.61\pm{0.05}$ & $5.83\pm{0.15}$ & $65\pm13$   &-&-&  49.21-49.78 \\
\#9$^\dag$  &   $2.99\pm0.06$ & O3-O7V & -  & $4.61\pm{0.05}$ & $5.58\pm{0.13}$ & $47\pm7$  &-&-& 48.98-49.51 \\
\#10  &  $2.83\pm0.29$$^\S$  & O5-O7V & -  & $4.59\pm{0.02}$ & $5.41\pm{0.42}$ & $43\pm17$ &-&-& 48.52-49.58\\ 
\#11  &  $2.96\pm0.09$  & O3-O5V & -  & $4.63\pm{0.02}$ & $5.57\pm{0.09}$ & $48\pm4$  &-&-& 49.23-49.46 \\ 
\#12  &  $3.00\pm0.08$  & O5-O7V & -  & $4.59\pm{0.02}$ & $5.45\pm{0.09}$ & $40\pm4$  &-&-& 48.89-49.29 \\ 
\#14  &  $2.98\pm0.03$ & O3-O5V & -  & $4.63\pm{0.02}$ & $5.53\pm{0.08}$ & $46\pm4$   &-&-& 49.20-49.41 \\ 
\#15 &   $4.52\pm0.16$$^\ddag$  & O3If*-O4If  &  O2-3.5If* & $4.64\pm{0.01}$$^\ddag$ & $6.11\pm(0.10)$$^\ddag$  & $96\pm14$$^\ddag$  &100&40& 49.85-49.90$^\ddag$ \\ 
\#17  &   $2.63\pm0.20$  & O7-O9.5V & -  & $4.54\pm{0.03}$ & $5.01\pm{0.13}$ & $24\pm3$  &-&-& 47.79-48.68 \\ 
\#18  & $3.21\pm0.16$ & O8-O9.5V & -  & $4.52\pm{0.02}$ & $5.20\pm{0.11}$  &  $28\pm3$  &-&-& 48.02-48.69  \\
\hline
\label{OBtable}                  
\end{tabular}
\tablefoot{
\tablefoottext{\dag}{Stars also observed by ISAAC/VLT (PI: N. Homeier, Program ID: 073.D-0837)} 
\tablefoottext{\ddag}{Parameters from photospheric model fitting}
\tablefoottext{\S}{The extinction of \#10 is taken from the value of nearest star \#1}
\tablefoottext{a} {Spectral types obtained by comparison with reference stars from \citet{Hanson:2005aa}} 
\tablefoottext{b} {Spectral types for Of stars according to criteria from \citet{Crowther:2011aa}}
\tablefoottext{c} {The upper and lower mass limits of \#1, \#2 and \#15  for the assumption of chemical homogeneous hydrogen and helium burners using the mass-luminosity relation from  \citet{Grafener:2011aa} (Sect. 3.4).}
}
\end{table*}

\subsubsection{Spectral classification of very massive stars}

\begin{figure*}
   \centering
   \includegraphics[width=0.9\hsize]{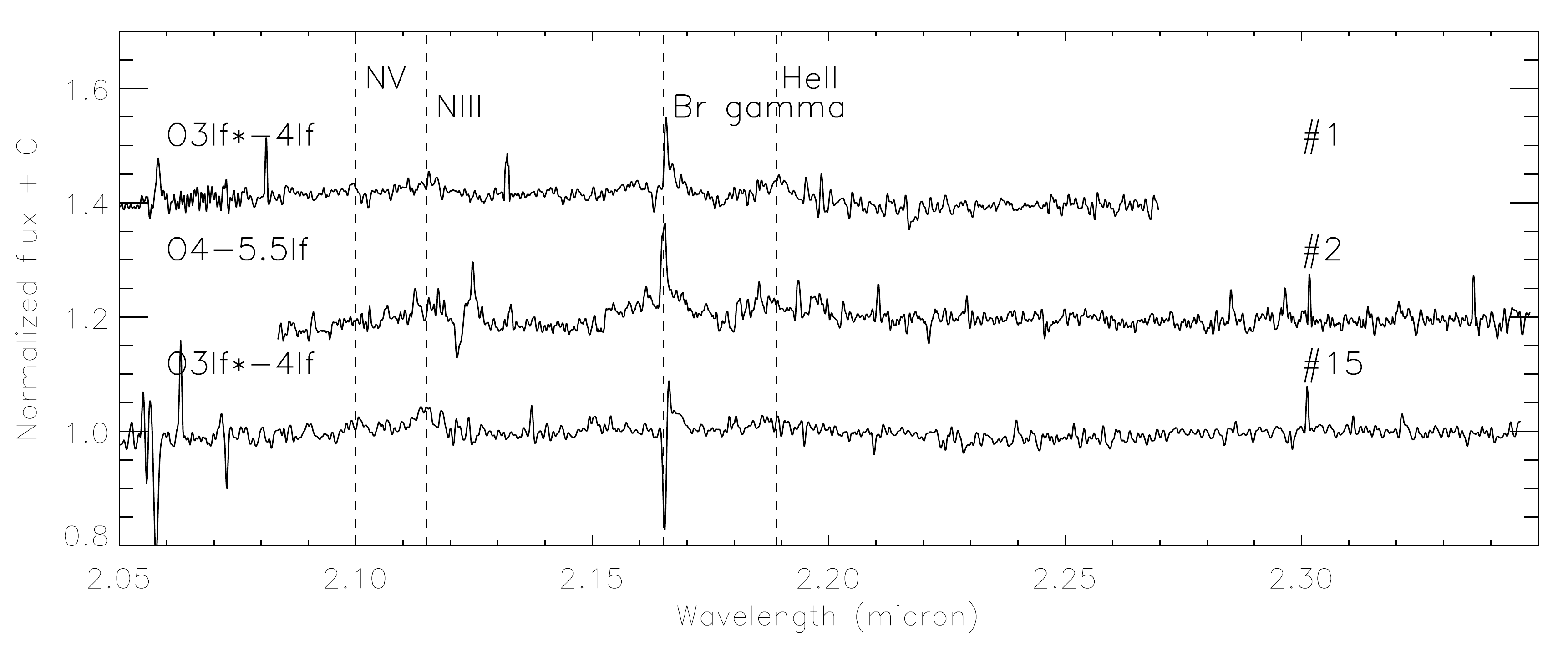}
   \caption{Normalized $K$-band spectra of Of type stars in W49 as taken with the multi-object mode of LUCI. The star number and the spectral type from spectroscopic classification are marked above the corresponding spectrum. Indicated with dashed lines are the spectral features crucial for classification.}
   \label{OfSpectra}
\end{figure*}

In addition to W49nr1 (\#1 in this paper) presented in \citet{Wu:2014aa} we have identified two additional stars (\#2 and \#15) whose spectra display broad $Br\gamma$ and \ion{He}{II} emission lines (Fig \ref{OfSpectra}). Additional lines of \ion{He}{I} and \ion{N}{III} are visible as well. We do not detect the \ion{N}{V} line at 2.10 \micron\ in \#2; it is only detected in \#1 and \#15. These spectral features are indicative of a strong stellar wind and a high effective temperature and we apply the same classification criteria as done for \#1 in \citet{Wu:2014aa}. 
 
The sum of the equivalent widths of their $Br\gamma$ and \ion{He}{II} lines (14.3 $\pm2.5$ \AA\ for \#2 and 4.4 $\pm1.2$ \AA\ for \#15, respectively) corresponds to a spectral type O2-3.5If* in the classification scheme proposed by \citet{Crowther:2011aa}, while the value for slash-stars (Of/WN) is in the range of $\sim 20-70$ \AA\ and lager than $\sim 70$ \AA\ for Wolf-Rayet stars. This criterion classifies \#2 as a slash-star with strong wind features. The absence of the \ion{N}{V} line in the spectrum of \#2 implies a later subtype and a lower temperature when compared to \#1 and \#15.

A comparison with reference stars with spectral types O3If*, O4If, O5If and O5.5If from \citet{Hanson:2005aa} shows the presence of the \ion{N}{V} line at 2.10 \micron\ in O3If* stars and some of the O4If stars. Accordingly \#2 is classified as a later type (O4If-O5.5If), while \#15 has an earlier type (O3If*-O4If). 
Even though the reference stars do not include the spectral type O2If*,
the temperature range obtained from the spectroscopic
analysis suggest that stars \#1 and \#15 can be as early as O2If*.

\subsection{Photometric vs Spectroscopic spectral type}
After the spectral classification of all the massive stars, 
we now can compare our photometric spectral types derived in Section 3.1 to those derived from the LUCI spectroscopy. 
Most of the O type stars have photometric spectral types similar to the spectral types derived from spectroscopy, like \#3, \#5, \#7, \#9, \#11, \#12 and, \#17. For the two B stars (\#6 and \#13),  however, we find a large discrepancy. The difference between the photometric and spectroscopic spectral type suggests that these stars are located at a closer distance than W49. In fact, their location in the CCD (Fig~\ref{CCD}) shows that they have a lower extinction than the other massive stars, consistent with a closer distance. Therefore we identify stars \#6 and \#13 as foreground stars. 

We identify 5 stars located above the reddening line of an O3 star in Fig~\ref{CMD}. Stars \#4 and \#21 are spectroscopically identified as YSO and have an infrared excess contributing to the H- and K-band flux. The VMS stars (\#1, \#2, \#15) are also located above the O3 line. This is partly because they are more luminous than O3V stars as they are super giant stars. Additionally, their infrared colors are different from main sequence stars due to the presence of a strong stellar wind (see below). 

Additionally, two stars classified as O3-O7 main sequence stars (\#3 and \#8) are brighter than expected for an O3V star. Both stars are located in the radio source CC. Several explanations could be given for this discrepancy. Due to limitations of the empirical spectral libraries, the earliest spectral type available for classification is O3V. Using stellar parameters taken from \citet[Table~1]{Martins:2005aa}, this spectral type corresponds to an effective temperature of 44,600 K. Because of this limitation, our O3 spectral classification includes also all stars hotter and brighter than O3V stars. Another possibility is that the radio source CC could be a different star forming complex in the foreground. Placing star \#3 on the O3V line would require a distance of 7.1 kpc. 

Additionally we compare our  spectral types of the 22 stars to the mid-infrared classification by \citet{Saral:2015aa}. We find 11 sources in common, their classifications are listed in Table \ref{SpectralType}. In general we find a reasonable agreement between our and Saral's classification. The mid-infrared photometry of  sources located in the small HII regions, \#2 in W49 South and \#8 in CC, are likely contaminated by the diffuse emission due to the large beam of Spitzer/IRAC. The higher the number in our source list, the fainter its K-band magnitude, making the contamination of diffuse emission more important. 

\section{Hertzsprung-Russell diagram}

\begin{figure}
   \centering
   \includegraphics[width=\hsize]{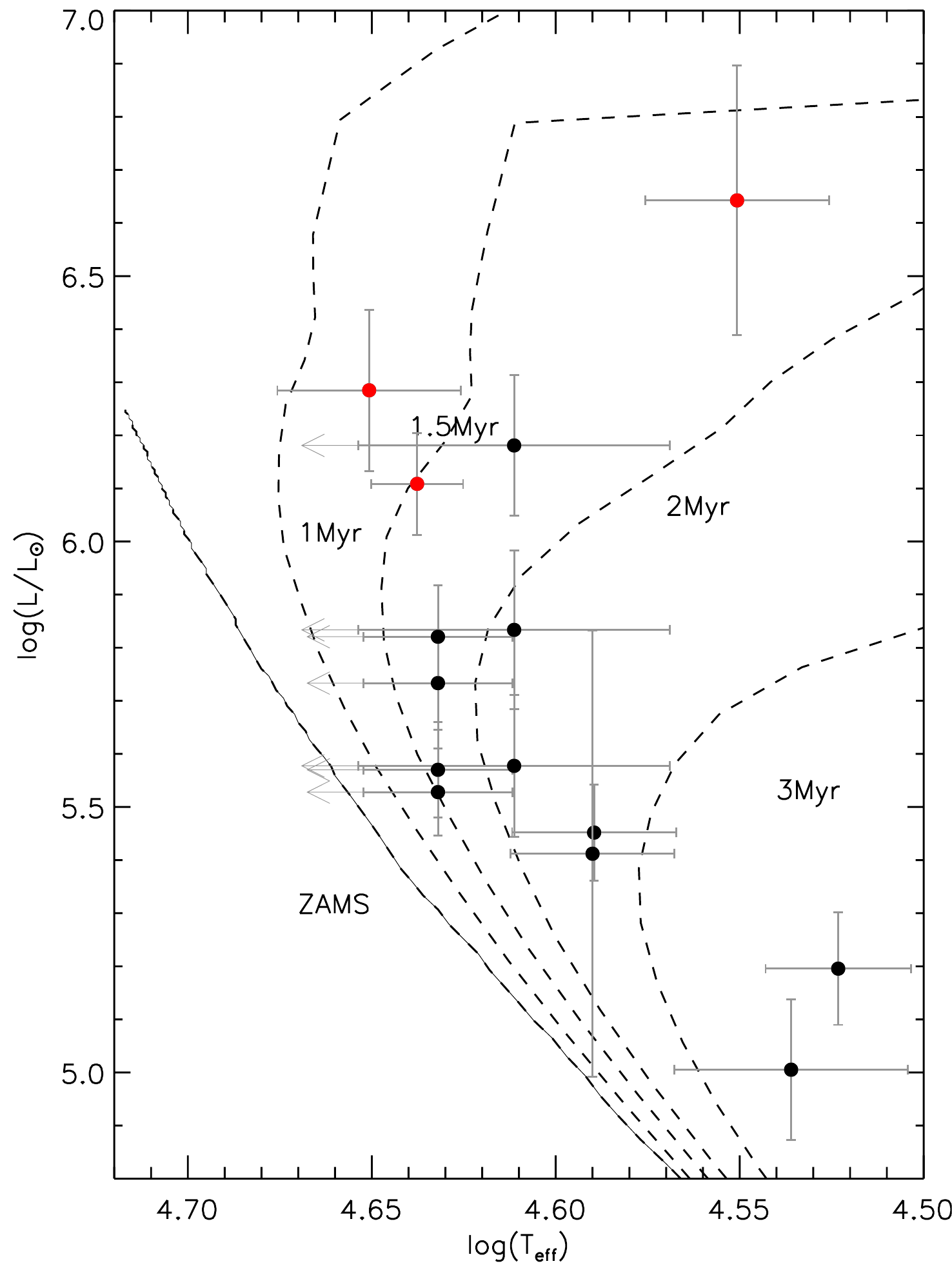}
      \caption{Hertzsprung-Russell diagram (HRD) of the massive stars in W49. The solid line represents the zero age main sequence isochrone from \citet{Lejeune:2001aa}. The dashed lines are main-sequence isochrones for 1, 1.5, 2 and 3 Myr from \citet{Ekstrom:2012aa} and \citet{Yusof:2013aa}. The stars are de-reddened using the extinction law of \citet{Indebetouw:2005aa}. Three very massive stars are indicated by red dots while OB main sequence stars are indicated by black dots.}
       \label{HRD}
\end{figure}

After the spectroscopic classification,  we are able to place the 11 OB stars and the 3 VMS stars in the HRD and we compare their locations with that of stellar evolution models. The effective temperature of O type main sequence stars is taken from \citet{Martins:2005aa}, while the bolometric correction and the intrinsic $H-K$ colors are from \citet{Martins:2006aa}. By assuming the \citet{Indebetouw:2005aa} extinction law, 
 $A_{K}$ was derived. For star \#10 which has only $K$-band photometry, the extinction is taken from the value of its nearest neighbour, \#1. With these parameters in hand, we can determine both the absolute bolometric magnitude and luminosity of the newly discovered massive stars  (Fig~\ref{HRD}). We use the stellar evolution tracks from the Geneva models  \citep{Ekstrom:2012aa,Yusof:2013aa} to derive the masses of the stars. The derived masses range from $\sim$20 \msun\ to $\sim$120 \msun\ for these stars. Those numbers as well as other parameters derived for the massive stars are compiled in Table~\ref{OBtable}.
 The large uncertainties in the locations of the stars are dominated by the uncertainty in the spectral classification.  The HRD is shown in Fig~\ref{HRD}, where the isochrones for 1 Myr, 1.5 Myr, 2 Myr and 3 Myr \citep{Ekstrom:2012aa,Yusof:2013aa} are plotted as dashed lines together with the Zero Age Main Sequence \citep[ZAMS,][]{Lejeune:2001aa}.

Due to the fact that O3 is the earliest spectral type available for classification, our O3 spectral classification includes also all stars hotter than 44,600 K. In the HRD for W49 (Fig~\ref{HRD}), stars with spectral types from O3 to O5 could be located at higher effective temperature than indicated, which would also affect their location with respect to the isochrones. A dedicated spectral modelling of the observed spectra is the way to overcome this limitation. 

To place \#1, \#2 and \#15 on the HRD, we derived $T_{eff}$ and
luminosity based on a grid of synthetic spectra computed with the non-LTE radiative transfer code {\sc cmfgen} \citep{Hillier:1998aa}. The stellar atmosphere models contain the following model atoms: \ion{H}{I}, \ion{He}{I-II}, \ion{C}{I-IV}, \ion{N}{I-V}, \ion{O}{I-VI}, \ion{Si}{II-IV} and \ion{Fe}{I-VII}. We set the surface gravity $\log g = 4$, the wind parameter $\beta = 1.0$, the volume filling factor $f_{\rm v} = 0.25$, the luminosity to $\log L/L_{\odot} = 6$ and the terminal velocity to typical values for O dwarfs. The effective temperature ($T_{eff}$) and the mass-loss rate ($\dot{M}$) were varied between 30,000 and 50,000K and $\log (\dot{M}/M_{\odot}) = -5$ and $-6.5$, respectively. The luminosity of our targets were estimated by extracting absolute magnitudes and intrinsic colors from the synthetic spectra.

From the best fitting
models we estimated $T_{eff}$ and extracted the intrinsic $H-K$
color. The intrinsic color was used to determine the extinction in the
$K$-band ($A_{K}$) by applying the extinction law of
\citet{Indebetouw:2005aa}. The absolute $K$-band magnitudes were
calculated by subtracting $A_{K}$ and the distance modulus from the
apparent magnitudes. The actual luminosity of the three stars was
obtained by rescaling the stellar model spectrum with a luminosity of
$\log L/L_{\odot} = 6$ to match the observed absolute magnitudes.

As a comparison to Geneva models, we apply the relation between the luminosity and the present-day stellar mass with the upper limit from homogeneous hydrogen burners and lower limit from helium burners \citep{Grafener:2011aa}. When adopting a hydrogen fraction of $0.7^{+0.05}_{-0.1}$, the upper mass limit would be in the range of  110-130 \msun\ for \#1, 200-240 \msun\ for \#2 and 90-100 \msun\ for \#15, in agreement with the masses estimated from tracks of massive star evolution. In case the three stars are helium burners, lower limits on their masses would be around 50 \msun, 90 \msun, and 40 \msun, respectively.

While the estimated upper mass limit of \#2 is considerably higher than the proposed upper mass limit of 150 \msun \citep{Figer:2005aa,Koen:2006aa}, the likelihood of very massive stars to be binary or multiple systems is also very high. X-ray observations can serve as diagnostics under the assumption that the intrinsic X-ray luminosity of single O stars can be approximated by $L_X/L_{Bol} \sim10^{-7}$ \citep{Chlebowski:1989aa,Crowther2010aa}. Colliding supersonic stellar winds in early type binaries will produce additional X-ray flux from the shock heated material in the wind interaction region \citep{Stevens:1992aa}. In our OB stars sample, only stars \#1, \#2 and \#12 are detected as bright X-ray sources by XMM and Chandra (Leisa Townsley, private communication) suggesting that they might be colliding-wind binaries. If this was the case, the masses of the individual components of these sources could be lower than the above estimate.

\begin{figure}
   \centering
   \includegraphics[width=\hsize]{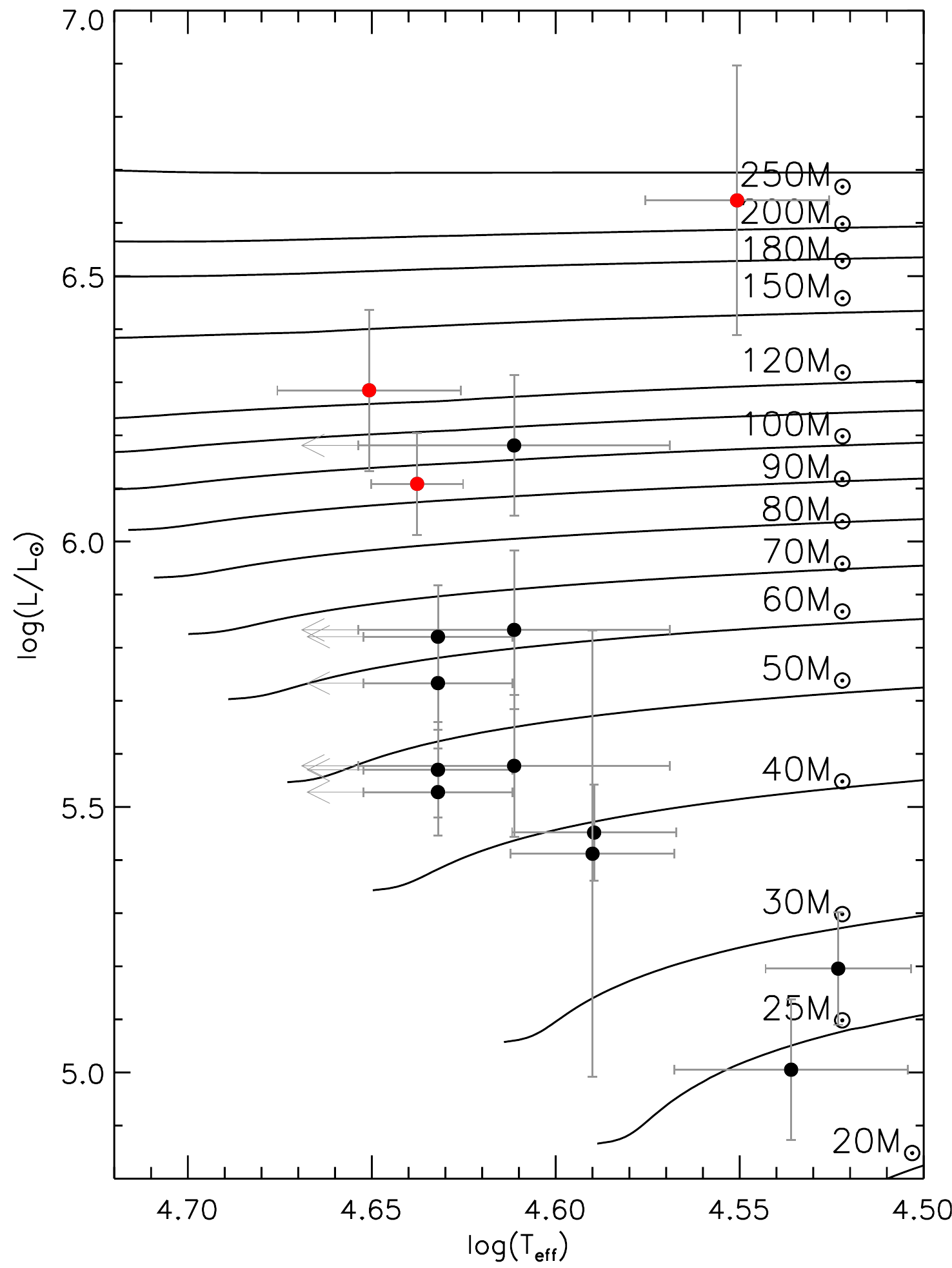}
      \caption{Hertzsprung-Russell diagram (HRD) of the massive stars in W49. The solid lines represent the evolutionary tracks from \citet{Ekstrom:2012aa} and \citet{Yusof:2013aa}. See Fig.~\ref{HRD} for an explanation of the symbols.}
       \label{HRDtracks}
\end{figure}

\section{Discussion}

\subsection{Cluster properties of W49}
\subsubsection{Cluster age}

On the HRD (Fig~\ref{HRD}), we overplotted the main sequence isochrones for different ages from the Geneva evolutionary model \citep{Ekstrom:2012aa,Yusof:2013aa} and compare the locations of the massive stars with the isochrones. Within the uncertainties, all stars (maybe with the exception of star \#18) are consistent with a 1.5 Myr isochrone. Smaller error bars would be required to make a definite statement about a possible age spread.

\subsubsection{Cluster mass}
From their photometric analysis, and by extrapolating the mass function to the mass range 1 to 20\,M$_\odot$, \citet{Homeier:2005aa} deduced a total stellar mass of 5 to 7$\times 10^4$ M$_\odot$ for a $5' \times 5'$ (16 $\times$ 16 pc) region centered on cluster 1. This estimate is limited by contamination with field stars, even after color selection. To quantify this contamination requires spectroscopy of every star or proper motion measurements to remove the fore- and background stars by their different spatial velocities \citep[e.g.][]{Stolte15}.

We can still try to place constraints on the shape of the IMF of W49 and determine whether the presence of 3 very massive stars (VMS) is consistent with the derived cluster mass by \citet{Homeier:2005aa} and a typical high-mass slope of -2.35 the IMF \citep{Salpeter55,Kroupa01}
We randomly sample 1000 times a \citet{Kroupa01} type mass function by drawing 200 high-mass stars between 20 and 400 \msun, corresponding to a total cluster mass between $5-7 \times 10^4$ \msun. We find that of these 200 stars, on average 10 $\pm$ 3 stars are expected to be more massive than 150 \msun. In W49 we have detected 3 VMS stars spectroscopically, suggesting that this detection is perfectly normal, and does not suggest any oddity in the mass function of W49. As we cannot quantify our spectroscopic completeness, we might have missed several VMS stars. The CMD (Fig~\ref{CMD}) shows several bright stars near the location of the spectroscopically detected VMS stars.

\subsection{Feedback}

Massive stars have a strong impact on their formation sites. The ionizing radiation and stellar winds of the massive stars are able to alter the state of the interstellar medium (ISM) and halt star formation or trigger new episodes. Observations of the ISM (e.g.\ \citet{Peng:2010aa,Galvan-Madrid:2013aa}) show the effect of the stellar feedback on the ISM. Simulations suggest that the hydrogen ionizing photons have a more destructive effect than the mechanical feedback of the stellar winds of massive stars \citep{Dale13}, however none of the two mechanisms is sufficient to fully destroy a giant molecular cloud of $\approx 10^6$\,M$_\odot$ surrounding a massive cluster.  

The energy and momentum input in the ISM can be quantified by a characterisation of the massive stars.  The spectral classification of the massive stars in W49 also results in an estimate of the number of ionizing photons ($Q_{0}$) emitted by each star and the stellar wind parameters. From \citet{Martins:2005aa}, we take the numbers of ionising photons emitted from stars with a certain spectral type. The ionising fluxes are rescaled with luminosities of individual stars and the final result listed in Table~\ref{OBtable} (\#1, \#2 and \#15 have the ionizing flux directly from photospheric modelling). The total amount of ionizing photons emitted in W49 is dominated by the 3 VMS ($10^{50.45\sim50.59}$ $s^{-1}$) and adds up to a total of ($10^{50.94} s^{-1}$) for the entire cluster (see Table~\ref{OBtable}). 

We can compare the number of ionizing photons with the radio flux emitted by W49. The radio free-free flux of the \ion{H}{II} region is a direct proxy of the number of ionizing photons and therefore can be compared with the stellar ionizing photons according to the spectral types we have derived. Low spatial resolution radio observations from  \citet{Kennicutt:1984aa}  provide us with the total integrated radio flux of the entire \ion{H}{II} region.  \citet{Kennicutt:1984aa}  derives a size of the radio emitting region of 60 pc, which includes all the massive stars we have identified. 

Based on their radio data, \citet{Kennicutt:1984aa} estimates the number of Lyman continuum photons to be  $10^{51.20} s^{-1}$ (adopting the distance of 14.1 kpc). When we scale this number to the currently most accurate distance of 11.1 kpc, the value decreases to $10^{50.99} s^{-1}$. The upper limit on the total number of ionizing photons emitted by our massive star sample is of the same order ($10^{50.94} s^{-1}$). 

High spatial resolution radio continuum observations of W49 \citep{de-Pree:1997aa} resolve the large \ion{H}{II} region detected by \citet{Kennicutt:1984aa} and reveal the presence of many smaller \ion{H}{II} regions which are harbouring one or more O stars. After comparing the spatial locations of the O stars with those of the \ion{H}{II} regions, we could identify two \ion{H}{II} regions where we have found a central O star: \#2 located in W49 South, and \#8 associated with the radio shell source CC \citep{de-Pree:1997aa}. We recalculated the number of ionizing photons in the two \ion{H}{II} regions based on \citet{de-Pree:1997aa}, and scaled to a distance of 11.1\,kpc. The Lyman continuum photon flux derived from radio observations of W49 South and CC are $10^{49.73} s^{-1}$ and $10^{48.96} s^{-1} $, respectively.

As listed in Table~\ref{OBtable}, the O stars \#2 and \#8 emit $10^{50.03\sim50.27}$ $s^{-1}$ and $10^{49.21\sim49.78}$ $s^{-1}$ Lyman continuum photons, respectively. In both cases, the early O type stars provide sufficient Lyman continuum flux to completely ionize the local \ion{H}{II} region, which is consistent with the assumption that the stars are the main, if not the only source of ionization in each region. 

In W49, the luminosity output from massive stars dominates the feedback towards the cloud. At the current stage, neither over-pressurized ionized gas nor radiation pressure from the central cluster have cleared the entire molecular cloud.  According to \citet{Galvan-Madrid:2013aa} only 1 \% of the total gas mass ($10^6$ \msun) is ionized. This is in agreement with simulations of cluster formation in massive molecular clouds \citep{Dale14}. Consequently, a large amount of photons must either be leaking out of W49 or must be absorbed by the dust still present in the region. Neither of these photons would contribute to the ionization of the nebula. We note that a large fraction of escaping ionizing photons is commonly observed in regions of massive star formation \citep{Kurtz94}.

Smaller scale effects of feedback inside the molecular cloud, however,  can be seen in W49. Fig.~\ref{W49JHK} shows the presence of a nebular arc north of the central cluster. This arc appears to be part of a ring structure seen in Spitzer images as well as CO line emission \citep{Peng:2010aa}. Rings are likely the result of the feedback by stellar winds or ionizing photons of a few massive stars in W49. 


%
%

\subsection{Do massive stars in W49 form in isolation?}

When looking at the spatial distribution of the massive stars  (Fig~\ref{W49JHK}) and the \ion{H}{II} regions (also tracing the locations of the more embedded O stars) it becomes clear that several star formation sites are present in W49. In the near-infrared a central cluster (cluster 1) becomes evident which contains "only" $10^4$ \msun\ within a radius of 45\arcsec (2.4 pc) \citep{Homeier:2005aa}, while the total mass of the 16 x 16 pc studied area is estimated to be $5-7 \times 10^4$ \msun. Apart from the clusters detected in the infrared, a proto-cluster is still forming and detected only at radio wavelength at a distance of  about 3 pc distance from the center of cluster 1.  The so called Welch ring \citep{Welch:1987aa} harbours dozens of strong radio continuum sources \citep{de-Pree:1997aa}, implying the presence of an even younger event of massive star formation. 

The majority of the massive stars are concentrated towards cluster 1. Within a radius of 45\arcsec (2.5 pc) we identify 10 OB stars and a massive YSO. \citet{Homeier:2005aa} used this radius to determine the total mass of cluster 1 to $10^4$ \msun\ . The four  other stars are located in two of the subclusters identified by \citet{Alves:2003aa}; one star in cluster 2 near the \ion{H}{II} region W49 South and three stars in cluster 3 associated with \ion{H}{II} region CC.

The environments of the two YSOs differ from each other. One of them (\#21) is in the subcluster to the northwest of the center (shown with cyan circles in Fig~\ref{W49JHK}). The other YSO (\#4) resides in the middle of the main cluster, thus showing that circumstellar disks can survive even in extreme environments, with high stellar density and strong UV radiation field.

The distribution of the 11 massive stars in cluster 1 shows that not all massive stars are located in a compact cluster -- unlike in NGC3603, where all the massive stars are within 1 pc of the center \citep{Moffat:1994aa}. Only 6 stars are within a projected distance less than 1 pc of the center of cluster 1, while 5 stars are located outside, including \#14 located at a projected distance of 2.2 pc.  The Spitzer images obtained with the IRAC camera \citep{Fazio04} suggest the presence of a bow shock to the north of star \#14  \citep{Saral:2015aa}. This would indicate that \#14 is a candidate run-away OB star, possibly orginating from the core of the central cluster  \citep{Gvaramadze:2010aa}. Assuming an age of 1 to 2 Myr (Sect. 4.1) we can calculate a projected velocity of 1 to 2 $km~s^{-1}$, which is required to reach 2.2 pc. This velocity is very low compared to the upper limit for OB runaway velocities of a few hundreds $km~s^{-1}$ \citep{Philp:1996aa}. Considering that bowshocks typically appear around stars with velocities of $>$ 10 $km~s^{-1}$, star \#14 might have been kicked out at a later state of the cluster lifetime. 

Using the above argument, also the massive stars associated with the W49 South and CC \ion{H}{II} regions might be considered to be runaway stars from the central cluster.  However, other arguments suggest that they might have formed there as members of a small sub cluster. The stars are inside an \ion{H}{II} region, still surrounded by the molecular material where they might have formed in. The shape of the W49 South \ion{H}{II} region \citep{de-Pree:1997aa} is cometary, but directed towards the center of cluster 1. The \ion{H}{II} region CC is classified as irregular, and thus unlike a cometary shape created by the interaction of moving O stars with an \ion{H}{II} region.  Additionally, \citet{Alves:2003aa} show that small sub clusters, consisting of more than one massive star are associated with W49 South and CC.

\begin{table*}
\caption{Physical parameters of star-forming regions}             
\centering          
\begin{tabular}{l c c c c c c}
\hline\hline       
Star-forming region &  Distance & Extinction  & Age & Cluster Mass  & Reference  \\ 
                          	&    (kpc)      &     (mag)   & (Myr) & ($10^4M_{\sun}$) & pc \\
\hline
W49  &  $11.1^{+0.8}_{-0.7}$  &  $A_{K_{\rm{s}}} \sim 3$  &  1-2  &  5-7  & 1, 2, 3, 4\\
NGC 3603  & $6.0\pm0.3$  & $A_{V} \sim 4.5$  &  1-2 &   $1.3\pm0.3$ & 5, 6, 7, 8 \\
W43  &  $5.49^{+0.39}_{-0.34}$  &  $A_V \sim 30$  &  5-10  &   -   &  9, 10, 11 \\
Westerlund 1 (Wd 1) &  $4.0\pm0.2$  & $A_{K_{\rm{s}}}=0.91\pm{0.05}$  &   3-6  &  5-15 & 7, 12, 13, 14, 15 \\
Westerlund 2 (Wd 2) &  $4.16\pm0.33$  &  $A_V=6.51\pm0.38$  &  < 3  &  >0.7 & 16, 17 \\
Carina Nebula Complex (CNC) & $2.3$ & $A_V=3.5$ & 2-8   &  >10 & 18, 19 \\
NGC 346 (in SMC) & $60.6$  &   -\tablefootmark{a}  & $\sim 3$ & 39 & 20, 21, 22 \\

\hline
\label{SFregions}                  
\end{tabular}
\tablefoot{
\tablefoottext{a} {The extinction towards NGC 346 is very low due to its location in the SMC; it is spatially highly variable and mainly comes from the star-forming region itself.} 
}
\tablebib{(1) \citet{Kennicutt:1984aa}; (2) \citet{Alves:2003aa}; (3) \citet{Homeier:2005aa}; (4) \citet{Zhang:2013ab};      
(5) \citet{Grabelsky:1988aa}; (6) \citet{Stolte:2004aa}; (7) \citet{Kudryavtseva:2012aa}; (8) \citet{Moffat:1994aa};          
(9) \citet{Zhang:2014aa}; (10) \citet{Blum:1999aa}; (11) \citet{Bally:2010aa};             
(12) \citet{Gennaro:2011aa}; (13) \citet{Clark:2005aa}; (14) \citet{Crowther:2006ab}; (15) \citet{Brandner:2008aa}          
(16) \citet{Vargas-Alvarez:2013aa}; (17) \citet{Ascenso:2007ab}    
(18) \citet{Preibisch:2012aa}; (19) \citet{Ascenso:2007aa}   
(20) \citet{Hilditch:2005aa}; (21) \citet{Bouret:2003aa}; (22) \citet{Sabbi:2008aa}. 
}
\end{table*}

\subsection{W49 as an extragalactic template?}

 The total gas mass within a radius of 60 pc and the stellar mass of W49 are estimated to be $~1.1\times10^6$ \msun \citep{Galvan-Madrid:2013aa} and $5-7\times10^4$ \msun \citep{Homeier:2005aa}, respectively. This classifies W49 as one of the most massive star-forming regions in the  Galaxy outside the Galactic center, and makes it a templates for extragalactic giant \ion{H}{II} regions, which have masses in the range $M_{\rm gas} = 10^4-10^7$\msun\ and  $M_{\rm stars} =  10^3-10^5$ \msun \citep{Kennicutt:1984aa}. W49 is reminiscent of the well-studied clusters NGC 3603 YC, W43, Westerlund 1 (Wd 1), Westerlund 2 (Wd 2) and the Carina Nebula Complex (CNC). Their physical parameters are shown in comparison with W49's in Table~\ref{SFregions}. The NGC 3603 YC \citep{Rochau} is very compact in the center and has $\sim$40 high-mass O- and WR stars confined into a very compact region of $\leq 1pc$ \citep{Moffat:1994aa}. The stellar cluster of Wd2 reveals a morphology similar to NGC 3603 YC, but the former's size is a few times larger and has a more extended scale. CNC consists of several dense clusters including Trumpler 14 and Trumpler 16, embedded in a large amount of gas and dust within the region extending over at least 80 pc \citep{Preibisch:2012aa}. The central region of W49 with massive stars confined into a single dense core shows a similar morphology to NGC 3603 \citep{Moffat:1994aa}, but its extent over more than 60 pc in diameter makes it less compact. Overall, the morphology of W49 is comparable to the Carina Nebula region, with some dense clusters embedded in a more distributed region of ongoing star formation, with the key difference that CNC is closer to us, and hence can be studied at higher spatial resolution \citep{Ascenso:2007aa}. 

The morphologies of star formation regions give us clues to understand their formation mechanisms. Different patterns of star formation exist among these regions. For NGC 3603, cloud-cloud collision might be the potential trigger of starburst \citep{Fukui:2014aa}, and the resulting monolithic collapse could explain the formation of a ultra-compact core of NGC 3603. This is probably not the case for W49, where the star formation event seems to be taking place over a larger volume with individual subclusterings. Star formation in W49 might have been triggered by the density waves of a spiral arm forming a giant molecular cloud, which subsequently via hierarchical fragmentation morphed into several cores of different mass and density, and finally resulting in cluster 1 and other subclusters in W49. For CNC, and very likely also for W49, star formation occurred not only in the core of the region, but originated independently in multiple cores away from the centre. Several young clusters formed and exist simultaneously in the Carina Nebula star formation complex. For W49, with several sub-clusters outside of the main cluster, the formation process seems not to be strictly coeval. The presence of the Welch ring also indicates non-coeval star formation. Considering the comparable size of W49 and CNC, and the presence of multi-seed star formation sites in W49, CNC with its several compact clusters may represent a more evolved state of a W49-type starburst region.

\section{Conclusions}
In this paper we present $JHK$ imaging (SOFI/NTT and LUCI/LBT) and $K$-band spectroscopy (LUCI/LBT) of the massive stellar content in W49, one of the most massive and young star-forming regions in our Galaxy located at a distance of 11.1 kpc from us. Our main findings are as follows.
   \begin{enumerate}
      \item Our photometry confirms the high extinction (on average $A_{K_{\rm{s}}} \sim 3$\,mag) as well as large extinction variations. The presence of two infrared excess sources implies the existence of circumstellar disks around massive YSOs.
      \item Fourteen O-type stars and two stars with CO disks are identified according to their NIR spectra. With the derived spectral types, their physical parameters, including effective temperature and bolometric luminosities are estimated. Along with the magnitudes derived from photometry and comparison with Geneva stellar evolution models, masses and ages of the massive stars are estimated. The most massive star found in our survey is \#2, with an upper mass limit of $\approx$240 \msun.
      \item The analysis of the stellar population enables us to study the properties of the cluster and the star formation region. Massive cluster members have a typical age of 1.5\,Myr, while the presence of embedded sources indicates still ongoing star formation in the region. The number of massive stars is consistent with previous estimates of a cluster mass of 50,000 to 70,000 solar masses. The massive stars are also capable of providing the vast majority of the ionizing photons powering the \ion{H}{II} region. The spatial distribution of the massive stars indicates that some of the stars might have formed several half-mass radii away from the cluster center, though ejection from the clusters is also a possibility.
      \item Considering its young age, abundant reservoir of gas, and high total mass, W49 is comparable to extragalactic giant \ion{H}{II} regions. It can serve as a template and help us to understand star formation in normal and starburst galaxies, which is poorly studied due to large distances.
      
   \end{enumerate}

Due to the extreme crowding in the cluster centre and the incompleteness of our spectroscopic survey sample, a complete stellar census of W49 is not possible at seeing limited angular resolution. This limits our ability to precisely reconstruct the formation history of W49. Higher angular resolution observation are required to achieve a comprehensive view of the formation of this young star cluster.

\begin{acknowledgements}
We thank the referee for helpful suggestions that improved the paper significantly. We acknowledge Boyke Rochau for carrying out part of the spectroscopic observations. We thank Nancy Ageorges and Walter Seifert for carrying out the K-band imaging observations. We thank Leisa Townsley and Patrick Broos for discussions on the X-ray sources in W49. A.B.\ acknowledges MPIA for hospitality and travel support.

IRAF is distributed by the National Optical Astronomy Observatory, which is operated by the Association of Universities for Research in Astronomy, Inc., under cooperative agreement with the National Science Foundation
\end{acknowledgements}

\bibliographystyle{aa}
\bibliography{my_bib,arjanW49_bib}

\begin{thebibliography}{90}
\expandafter\ifx\csname natexlab\endcsname\relax\def\natexlab#1{#1}\fi

\bibitem[{Ageorges {et~al.}(2010)Ageorges, Seifert, J{\"u}tte, Knierim,
  Lehmitz, Germeroth, Buschkamp, Polsterer, Pasquali, Naranjo, Gemperlein,
  Hill, Feiz, Hofmann, Laun, Lederer, Lenzen, Mall, Mandel, M{\"u}ller,
  Quirrenbach, Sch{\"a}ffner, Storz, \& Weiser}]{Ageorges10}
Ageorges, N., Seifert, W., J{\"u}tte, M., {et~al.} 2010, \procspie, 7735, 53

\bibitem[{{Alves} \& {Homeier}(2003)}]{Alves:2003aa}
{Alves}, J. \& {Homeier}, N. 2003, \apjl, 589, L45

\bibitem[{{Ascenso} {et~al.}(2007{\natexlab{a}}){Ascenso}, {Alves}, {Beletsky},
  \& {Lago}}]{Ascenso:2007ab}
{Ascenso}, J., {Alves}, J., {Beletsky}, Y., \& {Lago}, M.~T.~V.~T.
  2007{\natexlab{a}}, \aap, 466, 137

\bibitem[{{Ascenso} {et~al.}(2007{\natexlab{b}}){Ascenso}, {Alves}, {Vicente},
  \& {Lago}}]{Ascenso:2007aa}
{Ascenso}, J., {Alves}, J., {Vicente}, S., \& {Lago}, M.~T.~V.~T.
  2007{\natexlab{b}}, \aap, 476, 199

\bibitem[{{Bally} {et~al.}(2010){Bally}, {Anderson}, {Battersby}, {Calzoletti},
  {Digiorgio}, {Faustini}, {Ginsburg}, {Li}, {Nguyen Luong}, {Molinari},
  {Motte}, {Pestalozzi}, {Plume}, {Rodon}, {Schilke}, {Schlingman},
  {Schneider-Bontemps}, {Shirley}, {Stringfellow}, {Testi}, {Traficante},
  {Veneziani}, \& {Zavagno}}]{Bally:2010aa}
{Bally}, J., {Anderson}, L.~D., {Battersby}, C., {et~al.} 2010, \aap, 518, L90

\bibitem[{{Banerjee} {et~al.}(2012){Banerjee}, {Kroupa}, \&
  {Oh}}]{Banerjee:2012aa}
{Banerjee}, S., {Kroupa}, P., \& {Oh}, S. 2012, \apj, 746, 15

\bibitem[{{Bestenlehner} {et~al.}(2011){Bestenlehner}, {Vink}, {Gr{\"a}fener},
  {Najarro}, {Evans}, {Bastian}, {Bonanos}, {Bressert}, {Crowther}, {Doran},
  {Friedrich}, {H{\'e}nault-Brunet}, {Herrero}, {de Koter}, {Langer}, {Lennon},
  {Ma{\'{\i}}z Apell{\'a}niz}, {Sana}, {Soszynski}, \&
  {Taylor}}]{Bestenlehner:2011aa}
{Bestenlehner}, J.~M., {Vink}, J.~S., {Gr{\"a}fener}, G., {et~al.} 2011, \aap,
  530, L14

\bibitem[{{Bik} {et~al.}(2012){Bik}, {Henning}, {Stolte}, {Brandner},
  {Gouliermis}, {Gennaro}, {Pasquali}, {Rochau}, {Beuther}, {Ageorges},
  {Seifert}, {Wang}, \& {Kudryavtseva}}]{Bik:2012aa}
{Bik}, A., {Henning}, T., {Stolte}, A., {et~al.} 2012, \apj, 744, 87

\bibitem[{{Bik} {et~al.}(2005){Bik}, {Kaper}, {Hanson}, \&
  {Smits}}]{Bik:2005aa}
{Bik}, A., {Kaper}, L., {Hanson}, M.~M., \& {Smits}, M. 2005, \aap, 440, 121

\bibitem[{{Bik} {et~al.}(2006){Bik}, {Kaper}, \& {Waters}}]{Bik:2006aa}
{Bik}, A., {Kaper}, L., \& {Waters}, L.~B.~F.~M. 2006, \aap, 455, 561

\bibitem[{{Bik} {et~al.}(2014){Bik}, {Stolte}, {Gennaro}, {Brandner},
  {Gouliermis}, {Hu{\ss}mann}, {Tognelli}, {Rochau}, {Henning}, {Adamo},
  {Beuther}, {Pasquali}, \& {Wang}}]{Bik:2014aa}
{Bik}, A., {Stolte}, A., {Gennaro}, M., {et~al.} 2014, \aap, 561, A12

\bibitem[{{Bik} \& {Thi}(2004)}]{Bik:2004aa}
{Bik}, A. \& {Thi}, W.~F. 2004, \aap, 427, L13

\bibitem[{{Blaauw}(1991)}]{Blaauw:1991aa}
{Blaauw}, A. 1991, in NATO Advanced Science Institutes (ASI) Series C, Vol.
  342, NATO Advanced Science Institutes (ASI) Series C, ed. C.~J. {Lada} \&
  N.~D. {Kylafis}, 125

\bibitem[{{Blum} {et~al.}(1999){Blum}, {Damineli}, \& {Conti}}]{Blum:1999aa}
{Blum}, R.~D., {Damineli}, A., \& {Conti}, P.~S. 1999, \aj, 117, 1392

\bibitem[{{Bonnell} {et~al.}(2004){Bonnell}, {Vine}, \&
  {Bate}}]{Bonnell:2004aa}
{Bonnell}, I.~A., {Vine}, S.~G., \& {Bate}, M.~R. 2004, \mnras, 349, 735

\bibitem[{{Bouret} {et~al.}(2003){Bouret}, {Lanz}, {Hillier}, {Heap}, {Hubeny},
  {Lennon}, {Smith}, \& {Evans}}]{Bouret:2003aa}
{Bouret}, J.-C., {Lanz}, T., {Hillier}, D.~J., {et~al.} 2003, \apj, 595, 1182

\bibitem[{{Brandner} {et~al.}(2008){Brandner}, {Clark}, {Stolte}, {Waters},
  {Negueruela}, \& {Goodwin}}]{Brandner:2008aa}
{Brandner}, W., {Clark}, J.~S., {Stolte}, A., {et~al.} 2008, \aap, 478, 137

\bibitem[{{Bressert} {et~al.}(2012){Bressert}, {Bastian}, {Evans}, {Sana},
  {H{\'e}nault-Brunet}, {Goodwin}, {Parker}, {Gieles}, {Bestenlehner}, {Vink},
  {Taylor}, {Crowther}, {Longmore}, {Gr{\"a}fener}, {Ma{\'{\i}}z
  Apell{\'a}niz}, {de Koter}, {Cantiello}, \& {Kruijssen}}]{Bressert:2012aa}
{Bressert}, E., {Bastian}, N., {Evans}, C.~J., {et~al.} 2012, \aap, 542, A49

\bibitem[{Buschkamp {et~al.}(2010)Buschkamp, Hofmann, Gemperlein, Polsterer,
  Ageorges, Eisenhauer, Lederer, Honsberg, Haug, Eibl, Seifert, \&
  Genzel}]{Buschkamp10}
Buschkamp, P., Hofmann, R., Gemperlein, H., {et~al.} 2010, \procspie, 7735, 236

\bibitem[{{Chlebowski} {et~al.}(1989){Chlebowski}, {Harnden}, \&
  {Sciortino}}]{Chlebowski:1989aa}
{Chlebowski}, T., {Harnden}, Jr., F.~R., \& {Sciortino}, S. 1989, \apj, 341,
  427

\bibitem[{{Clark} {et~al.}(2005){Clark}, {Negueruela}, {Crowther}, \&
  {Goodwin}}]{Clark:2005aa}
{Clark}, J.~S., {Negueruela}, I., {Crowther}, P.~A., \& {Goodwin}, S.~P. 2005,
  \aap, 434, 949

\bibitem[{{Comer{\'o}n} \& {Pasquali}(2012)}]{Comeron:2012aa}
{Comer{\'o}n}, F. \& {Pasquali}, A. 2012, \aap, 543, A101

\bibitem[{Conti \& Blum(2002)}]{Conti02}
Conti, P.~S. \& Blum, R.~D. 2002, \apj, 564, 827

\bibitem[{{Crowther} {et~al.}(2006){Crowther}, {Hadfield}, {Clark},
  {Negueruela}, \& {Vacca}}]{Crowther:2006ab}
{Crowther}, P.~A., {Hadfield}, L.~J., {Clark}, J.~S., {Negueruela}, I., \&
  {Vacca}, W.~D. 2006, \mnras, 372, 1407

\bibitem[{{Crowther} {et~al.}(2010){Crowther}, {Schnurr}, {Hirschi}, {Yusof},
  {Parker}, {Goodwin}, \& {Kassim}}]{Crowther2010aa}
{Crowther}, P.~A., {Schnurr}, O., {Hirschi}, R., {et~al.} 2010, \mnras, 408,
  731

\bibitem[{{Crowther} \& {Walborn}(2011)}]{Crowther:2011aa}
{Crowther}, P.~A. \& {Walborn}, N.~R. 2011, \mnras, 416, 1311

\bibitem[{Dale {et~al.}(2013)Dale, Ngoumou, Ercolano, \& Bonnell}]{Dale13}
Dale, J.~E., Ngoumou, J., Ercolano, B., \& Bonnell, I.~A. 2013, \mnras, 436,
  3430

\bibitem[{Dale {et~al.}(2014)Dale, Ngoumou, Ercolano, \& Bonnell}]{Dale14}
Dale, J.~E., Ngoumou, J., Ercolano, B., \& Bonnell, I.~A. 2014, \mnras, 442,
  694

\bibitem[{{Davies}(2007)}]{Davies:2007aa}
{Davies}, R.~I. 2007, \mnras, 375, 1099

\bibitem[{{de Pree} {et~al.}(1997){de Pree}, {Mehringer}, \&
  {Goss}}]{de-Pree:1997aa}
{de Pree}, C.~G., {Mehringer}, D.~M., \& {Goss}, W.~M. 1997, \apj, 482, 307

\bibitem[{{de Wit} {et~al.}(2005){de Wit}, {Testi}, {Palla}, \&
  {Zinnecker}}]{de-Wit:2005aa}
{de Wit}, W.~J., {Testi}, L., {Palla}, F., \& {Zinnecker}, H. 2005, \aap, 437,
  247

\bibitem[{{de Zeeuw} {et~al.}(1999){de Zeeuw}, {Hoogerwerf}, {de Bruijne},
  {Brown}, \& {Blaauw}}]{de-Zeeuw:1999aa}
{de Zeeuw}, P.~T., {Hoogerwerf}, R., {de Bruijne}, J.~H.~J., {Brown}, A.~G.~A.,
  \& {Blaauw}, A. 1999, \aj, 117, 354

\bibitem[{{Ekstr{\"o}m} {et~al.}(2012){Ekstr{\"o}m}, {Georgy}, {Eggenberger},
  {Meynet}, {Mowlavi}, {Wyttenbach}, {Granada}, {Decressin}, {Hirschi},
  {Frischknecht}, {Charbonnel}, \& {Maeder}}]{Ekstrom:2012aa}
{Ekstr{\"o}m}, S., {Georgy}, C., {Eggenberger}, P., {et~al.} 2012, \aap, 537,
  A146

\bibitem[{Fang {et~al.}(2012)Fang, van Boekel, King, Henning, Bouwman, Doi,
  Okamoto, Roccatagliata, \& Sicilia-Aguilar}]{Fang12}
Fang, M., van Boekel, R., King, R.~R., {et~al.} 2012, \aap, 539, A119

\bibitem[{Fazio {et~al.}(2004)Fazio, Hora, Allen, Ashby, Barmby, Deutsch,
  Huang, Kleiner, Marengo, Megeath, Melnick, Pahre, Patten, Polizotti, Smith,
  Taylor, Wang, Willner, Hoffmann, Pipher, Forrest, McMurty, McCreight,
  McKelvey, McMurray, Koch, Moseley, Arendt, Mentzell, Marx, Losch, Mayman,
  Eichhorn, Krebs, Jhabvala, Gezari, Fixsen, Flores, Shakoorzadeh, Jungo,
  Hakun, Workman, Karpati, Kichak, Whitley, Mann, Tollestrup, Eisenhardt,
  Stern, Gorjian, Bhattacharya, Carey, Nelson, Glaccum, Lacy, Lowrance, Laine,
  Reach, Stauffer, Surace, Wilson, Wright, Hoffman, Domingo, \&
  Cohen}]{Fazio04}
Fazio, G.~G., Hora, J.~L., Allen, L.~E., {et~al.} 2004, \apjs, 154, 10

\bibitem[{{Figer}(2005)}]{Figer:2005aa}
{Figer}, D.~F. 2005, \nat, 434, 192

\bibitem[{Fitzpatrick(1999)}]{Fitzpatrick99}
Fitzpatrick, E.~L. 1999, \pasp, 111, 63

\bibitem[{{Fukui} {et~al.}(2014){Fukui}, {Ohama}, {Hanaoka}, {Furukawa},
  {Torii}, {Dawson}, {Mizuno}, {Hasegawa}, {Fukuda}, {Soga}, {Moribe},
  {Kuroda}, {Hayakawa}, {Kawamura}, {Kuwahara}, {Yamamoto}, {Okuda}, {Onishi},
  {Maezawa}, \& {Mizuno}}]{Fukui:2014aa}
{Fukui}, Y., {Ohama}, A., {Hanaoka}, N., {et~al.} 2014, \apj, 780, 36

\bibitem[{{Galvan-Madrid} {et~al.}(2013){Galvan-Madrid}, {Liu}, {Zhang},
  {Pineda}, {Peng}, {Zhang}, {Keto}, {Ho}, {Rodriguez}, {Zapata}, {Peters}, {De
  Pree}, \& {.}}]{Galvan-Madrid:2013aa}
{Galvan-Madrid}, R., {Liu}, H.~B., {Zhang}, Z.-Y., {et~al.} 2013, ArXiv
  e-prints

\bibitem[{{Gennaro} {et~al.}(2011){Gennaro}, {Brandner}, {Stolte}, \&
  {Henning}}]{Gennaro:2011aa}
{Gennaro}, M., {Brandner}, W., {Stolte}, A., \& {Henning}, T. 2011, \mnras,
  412, 2469

\bibitem[{{Grabelsky} {et~al.}(1988){Grabelsky}, {Cohen}, {Bronfman}, \&
  {Thaddeus}}]{Grabelsky:1988aa}
{Grabelsky}, D.~A., {Cohen}, R.~S., {Bronfman}, L., \& {Thaddeus}, P. 1988,
  \apj, 331, 181

\bibitem[{{Gr{\"a}fener} {et~al.}(2011){Gr{\"a}fener}, {Vink}, {de Koter}, \&
  {Langer}}]{Grafener:2011aa}
{Gr{\"a}fener}, G., {Vink}, J.~S., {de Koter}, A., \& {Langer}, N. 2011, \aap,
  535, A56

\bibitem[{{Gvaramadze} {et~al.}(2010){Gvaramadze}, {Kroupa}, \&
  {Pflamm-Altenburg}}]{Gvaramadze:2010aa}
{Gvaramadze}, V.~V., {Kroupa}, P., \& {Pflamm-Altenburg}, J. 2010, \aap, 519,
  A33

\bibitem[{{Hanson} {et~al.}(1996){Hanson}, {Conti}, \& {Rieke}}]{Hanson:1996aa}
{Hanson}, M.~M., {Conti}, P.~S., \& {Rieke}, M.~J. 1996, \apjs, 107, 281

\bibitem[{{Hanson} {et~al.}(2005){Hanson}, {Kudritzki}, {Kenworthy}, {Puls}, \&
  {Tokunaga}}]{Hanson:2005aa}
{Hanson}, M.~M., {Kudritzki}, R.-P., {Kenworthy}, M.~A., {Puls}, J., \&
  {Tokunaga}, A.~T. 2005, \apjs, 161, 154

\bibitem[{Hern{\'a}ndez {et~al.}(2008)Hern{\'a}ndez, Hartmann, Calvet,
  Jeffries, Gutermuth, Muzerolle, \& Stauffer}]{Hernandez08}
Hern{\'a}ndez, J., Hartmann, L., Calvet, N., {et~al.} 2008, \apj, 686, 1195

\bibitem[{{Hilditch} {et~al.}(2005){Hilditch}, {Howarth}, \&
  {Harries}}]{Hilditch:2005aa}
{Hilditch}, R.~W., {Howarth}, I.~D., \& {Harries}, T.~J. 2005, \mnras, 357, 304

\bibitem[{Hill {et~al.}(2006)Hill, Green, \& Slagle}]{Hill06}
Hill, J.~M., Green, R.~F., \& Slagle, J.~H. 2006, \procspie, 6267, 31

\bibitem[{{Hillier} \& {Miller}(1998)}]{Hillier:1998aa}
{Hillier}, D.~J. \& {Miller}, D.~L. 1998, \apj, 496, 407

\bibitem[{{Hollenbach} {et~al.}(2000){Hollenbach}, {Yorke}, \&
  {Johnstone}}]{Hollenbach:2000aa}
{Hollenbach}, D.~J., {Yorke}, H.~W., \& {Johnstone}, D. 2000, Protostars and
  Planets IV, 401

\bibitem[{{Homeier} \& {Alves}(2005)}]{Homeier:2005aa}
{Homeier}, N.~L. \& {Alves}, J. 2005, \aap, 430, 481

\bibitem[{{Indebetouw} {et~al.}(2005){Indebetouw}, {Mathis}, {Babler}, {Meade},
  {Watson}, {Whitney}, {Wolff}, {Wolfire}, {Cohen}, {Bania}, {Benjamin},
  {Clemens}, {Dickey}, {Jackson}, {Kobulnicky}, {Marston}, {Mercer},
  {Stauffer}, {Stolovy}, \& {Churchwell}}]{Indebetouw:2005aa}
{Indebetouw}, R., {Mathis}, J.~S., {Babler}, B.~L., {et~al.} 2005, \apj, 619,
  931

\bibitem[{{Johnston} {et~al.}(2014){Johnston}, {Beuther}, {Linz}, {Schmiedeke},
  {Ragan}, \& {Henning}}]{Johnston:2014aa}
{Johnston}, K.~G., {Beuther}, H., {Linz}, H., {et~al.} 2014, \aap, 568, A56

\bibitem[{{Kennicutt}(1984)}]{Kennicutt:1984aa}
{Kennicutt}, Jr., R.~C. 1984, \apj, 287, 116

\bibitem[{{Koen}(2006)}]{Koen:2006aa}
{Koen}, C. 2006, \mnras, 365, 590

\bibitem[{Kroupa(2001)}]{Kroupa01}
Kroupa, P. 2001, \mnras, 322, 231

\bibitem[{{Kudryavtseva} {et~al.}(2012){Kudryavtseva}, {Brandner}, {Gennaro},
  {Rochau}, {Stolte}, {Andersen}, {Da Rio}, {Henning}, {Tognelli}, {Hogg},
  {Clark}, \& {Waters}}]{Kudryavtseva:2012aa}
{Kudryavtseva}, N., {Brandner}, W., {Gennaro}, M., {et~al.} 2012, \apjl, 750,
  L44

\bibitem[{Kurtz {et~al.}(1994)Kurtz, Churchwell, \& Wood}]{Kurtz94}
Kurtz, S.~E., Churchwell, E.~B., \& Wood, D. O.~S. 1994, \apjs, 91, 659

\bibitem[{{Lejeune} \& {Schaerer}(2001)}]{Lejeune:2001aa}
{Lejeune}, T. \& {Schaerer}, D. 2001, \aap, 366, 538

\bibitem[{{Longmore} {et~al.}(2014){Longmore}, {Kruijssen}, {Bastian}, {Bally},
  {Rathborne}, {Testi}, {Stolte}, {Dale}, {Bressert}, \&
  {Alves}}]{Longmore:2014aa}
{Longmore}, S.~N., {Kruijssen}, J.~M.~D., {Bastian}, N., {et~al.} 2014,
  Protostars and Planets VI, 291

\bibitem[{{Martins} \& {Plez}(2006)}]{Martins:2006aa}
{Martins}, F. \& {Plez}, B. 2006, \aap, 457, 637

\bibitem[{{Martins} {et~al.}(2005){Martins}, {Schaerer}, \&
  {Hillier}}]{Martins:2005aa}
{Martins}, F., {Schaerer}, D., \& {Hillier}, D.~J. 2005, \aap, 436, 1049

\bibitem[{McGregor {et~al.}(1988)McGregor, Hyland, \& Hillier}]{Mcgregor88}
McGregor, P.~J., Hyland, A.~R., \& Hillier, D.~J. 1988, \apj, 324, 1071

\bibitem[{{Moffat} {et~al.}(1994){Moffat}, {Drissen}, \&
  {Shara}}]{Moffat:1994aa}
{Moffat}, A.~F.~J., {Drissen}, L., \& {Shara}, M.~M. 1994, \apj, 436, 183

\bibitem[{Nagy {et~al.}(2015)Nagy, van~der Tak, Fuller, \& Plume}]{Nagy15}
Nagy, Z., van~der Tak, F. F.~S., Fuller, G.~A., \& Plume, R. 2015, \aap, 577,
  127

\bibitem[{Nagy {et~al.}(2012)Nagy, van~der Tak, Fuller, Spaans, \&
  Plume}]{Nagy12}
Nagy, Z., van~der Tak, F. F.~S., Fuller, G.~A., Spaans, M., \& Plume, R. 2012,
  \aap, 542, 6

\bibitem[{Nishiyama {et~al.}(2009)Nishiyama, Tamura, Hatano, Kato, Tanab{\'e},
  Sugitani, \& Nagata}]{Nishiyama09}
Nishiyama, S., Tamura, M., Hatano, H., {et~al.} 2009, \apj, 696, 1407

\bibitem[{{Olczak} {et~al.}(2010){Olczak}, {Pfalzner}, \&
  {Eckart}}]{Olczak:2010aa}
{Olczak}, C., {Pfalzner}, S., \& {Eckart}, A. 2010, \aap, 509, A63

\bibitem[{{Peng} {et~al.}(2010){Peng}, {Wyrowski}, {van der Tak}, {Menten}, \&
  {Walmsley}}]{Peng:2010aa}
{Peng}, T.-C., {Wyrowski}, F., {van der Tak}, F.~F.~S., {Menten}, K.~M., \&
  {Walmsley}, C.~M. 2010, \aap, 520, A84

\bibitem[{{Philp} {et~al.}(1996){Philp}, {Evans}, {Leonard}, \&
  {Frail}}]{Philp:1996aa}
{Philp}, C.~J., {Evans}, C.~R., {Leonard}, P.~J.~T., \& {Frail}, D.~A. 1996,
  \aj, 111, 1220

\bibitem[{{Preibisch} {et~al.}(2012){Preibisch}, {Roccatagliata}, {Gaczkowski},
  \& {Ratzka}}]{Preibisch:2012aa}
{Preibisch}, T., {Roccatagliata}, V., {Gaczkowski}, B., \& {Ratzka}, T. 2012,
  \aap, 541, A132

\bibitem[{Rieke \& Lebofsky(1985)}]{Rieke85}
Rieke, G.~H. \& Lebofsky, M.~J. 1985, \apj, 288, 618

\bibitem[{{Roberts} {et~al.}(2011){Roberts}, {van der Tak}, {Fuller}, {Plume},
  \& {Bayet}}]{Roberts:2011aa}
{Roberts}, H., {van der Tak}, F.~F.~S., {Fuller}, G.~A., {Plume}, R., \&
  {Bayet}, E. 2011, \aap, 525, A107

\bibitem[{{Rochau} {et~al.}(2010){Rochau}, {Brandner}, {Stolte}, {Gennaro},
  {Gouliermis}, {Da Rio}, {Dzyurkevich}, \& {Henning}}]{Rochau}
{Rochau}, B., {Brandner}, W., {Stolte}, A., {et~al.} 2010, \apjl, 716, L90

\bibitem[{{Sabbi} {et~al.}(2008){Sabbi}, {Sirianni}, {Nota}, {Tosi},
  {Gallagher}, {Smith}, {Angeretti}, {Meixner}, {Oey}, {Walterbos}, \&
  {Pasquali}}]{Sabbi:2008aa}
{Sabbi}, E., {Sirianni}, M., {Nota}, A., {et~al.} 2008, \aj, 135, 173

\bibitem[{Salpeter(1955)}]{Salpeter55}
Salpeter, E.~E. 1955, \apj, 121, 161

\bibitem[{{Saral} {et~al.}(2015){Saral}, {Hora}, {Willis}, {Koenig},
  {Gutermuth}, \& {Talat Saygac}}]{Saral:2015aa}
{Saral}, G., {Hora}, J.~L., {Willis}, S.~E., {et~al.} 2015, ArXiv e-prints

\bibitem[{Seifert {et~al.}(2010)Seifert, Ageorges, Lehmitz, Buschkamp, Knierim,
  Polsterer, Germeroth, Pasquali, Naranjo, J{\"u}tte, Feiz, Gemperlein,
  Hofmann, Laun, Lederer, Lenzen, Mall, Mandel, M{\"u}ller, Quirrenbach,
  Sch{\"a}ffner, Storz, \& Weiser}]{Seifert10}
Seifert, W., Ageorges, N., Lehmitz, M., {et~al.} 2010, \procspie, 7735, 256

\bibitem[{{Stevens} {et~al.}(1992){Stevens}, {Blondin}, \&
  {Pollock}}]{Stevens:1992aa}
{Stevens}, I.~R., {Blondin}, J.~M., \& {Pollock}, A.~M.~T. 1992, \apj, 386, 265

\bibitem[{{Stolte} {et~al.}(2004){Stolte}, {Brandner}, {Brandl}, {Zinnecker},
  \& {Grebel}}]{Stolte:2004aa}
{Stolte}, A., {Brandner}, W., {Brandl}, B., {Zinnecker}, H., \& {Grebel}, E.~K.
  2004, \aj, 128, 765

\bibitem[{Stolte {et~al.}(2015)Stolte, Hussmann, Olczak, Brandner, Habibi,
  Ghez, Morris, Lu, Clarkson, \& Anderson}]{Stolte15}
Stolte, A., Hussmann, B., Olczak, C., {et~al.} 2015, eprint arXiv

\bibitem[{Stolte {et~al.}(2010)Stolte, Morris, Ghez, Do, Lu, Wright, Ballard,
  Mills, \& Matthews}]{Stolte10}
Stolte, A., Morris, M.~R., Ghez, A.~M., {et~al.} 2010, \apj, 718, 810

\bibitem[{{Vargas {\'A}lvarez} {et~al.}(2013){Vargas {\'A}lvarez},
  {Kobulnicky}, {Bradley}, {Kannappan}, {Norris}, {Cool}, \&
  {Miller}}]{Vargas-Alvarez:2013aa}
{Vargas {\'A}lvarez}, C.~A., {Kobulnicky}, H.~A., {Bradley}, D.~R., {et~al.}
  2013, \aj, 145, 125

\bibitem[{{Welch} {et~al.}(1987){Welch}, {Dreher}, {Jackson}, {Terebey}, \&
  {Vogel}}]{Welch:1987aa}
{Welch}, W.~J., {Dreher}, J.~W., {Jackson}, J.~M., {Terebey}, S., \& {Vogel},
  S.~N. 1987, Science, 238, 1550

\bibitem[{Wheelwright {et~al.}(2010)Wheelwright, Oudmaijer, de~Wit, Hoare,
  Lumsden, \& Urquhart}]{Wheelwright10}
Wheelwright, H.~E., Oudmaijer, R.~D., de~Wit, W.-J., {et~al.} 2010, \mnras,
  408, 1840

\bibitem[{{Wu} {et~al.}(2014){Wu}, {Bik}, {Henning}, {Pasquali}, {Brandner}, \&
  {Stolte}}]{Wu:2014aa}
{Wu}, S.-W., {Bik}, A., {Henning}, T., {et~al.} 2014, \aap, 568, L13

\bibitem[{{Yusof} {et~al.}(2013){Yusof}, {Hirschi}, {Meynet}, {Crowther},
  {Ekstr{\"o}m}, {Frischknecht}, {Georgy}, {Abu Kassim}, \&
  {Schnurr}}]{Yusof:2013aa}
{Yusof}, N., {Hirschi}, R., {Meynet}, G., {et~al.} 2013, \mnras, 433, 1114

\bibitem[{{Zhang} {et~al.}(2014){Zhang}, {Moscadelli}, {Sato}, {Reid},
  {Menten}, {Zheng}, {Brunthaler}, {Dame}, {Xu}, \& {Immer}}]{Zhang:2014aa}
{Zhang}, B., {Moscadelli}, L., {Sato}, M., {et~al.} 2014, \apj, 781, 89

\bibitem[{{Zhang} {et~al.}(2013){Zhang}, {Reid}, {Menten}, {Zheng},
  {Brunthaler}, {Dame}, \& {Xu}}]{Zhang:2013ab}
{Zhang}, B., {Reid}, M.~J., {Menten}, K.~M., {et~al.} 2013, \apj, 775, 79

\bibitem[{{Zinnecker} \& {Yorke}(2007)}]{Zinnecker:2007aa}
{Zinnecker}, H. \& {Yorke}, H.~W. 2007, \araa, 45, 481

\end{thebibliography}

\appendix
\section{Observing log}

\begin{sidewaystable*}
\caption{Spectroscopically observed massive stars in W49}             
\label{table:1}      
\centering          
\begin{tabular}{ll c c c c c l l}
\hline\hline       
Source & ID & RA(J2000) & Dec (J2000) &$J$ & $H$ & $K$ & Class & Date\\ 
	&		&(h m s) & ($^{\circ}$ $'$ $''$)	&(mag) &(mag)&(mag)& & yyyy-mm-dd \\
\hline                    
1&MOS08Klt135(VMS)$^\dag$ 	&19 10 17.4	&+09 06 21 	& $16.57\pm{0.18} $ & $13.47\pm{0.12}$ & $11.93\pm{0.10}$ & Of & 2010-05-14, 2004-08-06$^\dag$\\
2&MOS07Klt135 				&19 10 21.8	&+09 05 03	& $19.57\pm{0.14} $ & $14.89\pm{0.06}$  &  $12.29\pm{0.29}$  &Of & 2010-05-14 \\ 
3&MOS13Klt135 				&19 10 11.9      &+09 06 58& $17.62\pm{0.01} $ & $14.42\pm{0.02}$  &  $12.59\pm{0.01}$  & OB & 2010-05-14\\ 
4&MOS12Klt135 				&19 10 16.0      &+09 06 14& $17.99\pm{0.07} $ & $15.15\pm{0.05}$ & $12.69\pm{0.01}$ & YSO & 2010-05-14\\
5&MOS13Klt1407$^\dag$    		&19 10 17.7      &+09 06 21 & $17.13\pm{0.02} $ & $14.25\pm{0.05}$  &  $12.86\pm{0.05}$  & OB & 2011-04-11, 2011-05-13 \\
  & 							& 			& 		   & 			           &  				    & 				     &	     & 2004-08-06$^\dag$, 2004-08-04$^\dag$\\
6&MOS20Klt135 				&19 10 14.9      &+09 06 49& $16.49\pm0.01$ & $14.31\pm0.02$  &  $13.16\pm0.01$ & foreground & 2010-05-14\\ 
7&MOS11Klt135$^\dag$ 			&19 10 16.5     &+09 06 03& $17.30\pm{0.05} $ & $14.60\pm{0.02}$  &  $13.17\pm{0.04}$  & OB & 2010-05-14, 2004-08-07$^\dag$ \\ 
8&MOS07Klt1412$^\dag$    		&19 10 11.6  &+09 07 06& $18.81\pm{0.07} $ & $15.36\pm{0.07}$  &  $13.51\pm{0.06}$  & OB & 2011-04-11, 2011-05-13\\
  & 							& 			& 		   & 			           &  				    & 				     &	     & 2004-08-04$^\dag$, 2004-08-05$^\dag$\\
9&MOS07Klt145m2r$^\dag$ 		&19 10 16.6 & +09 06 19& $17.96\pm{0.02}$ & $15.15\pm{0.02}$ & $13.60\pm{0.02}$ &OB & 2011-05-11, 2011-05-13, 2004-08-06$^\dag$\\
10&MOS08Klt145m2 			&19 10 17.3 & +09 06 19& - & -  & $13.69\pm{0.99}$ &OB  & 2011-05-11, 2011-05-13\\ 
11&MOS16Klt145				&19 10 18.7  &+09 05 56& $18.22\pm{0.10} $ & $15.28\pm{0.02}$  &  $13.74\pm{0.05}$  &  OB  & 2011-05-08\\ 
12&MOS14Klt145 				&19 10 16.9 & +09 06 10& $18.35\pm{0.07} $ & $15.32\pm{0.04}$  &  $13.77\pm{0.02}$  & OB & 2011-05-08 \\ 
13&MOS12Klt145m2r 			&19 10 12.7  &+09 05 25& $17.24\pm0.02$ & $14.94\pm0.03$ & $13.87\pm0.01$ & foreground  & 2011-05-11, 2011-05-13\\
14&MOS10Klt1411 				&19 10 17.4 & +09 07 01& $18.20\pm{0.03} $ & $15.41\pm{0.01}$  &  $13.87\pm{0.01}$  & OB & 2011-04-11, 2011-05-13\\ 
15&MOS12Klt15				&19 10 19.0 	&+09 06 23  	& - & $16.46\pm{0.08}$  &  $14.01\pm{0.05}$  & Of & 2011-05-14, 2011-05-15 \\ 
16&MOS19OBS45left 			&19 10 17.5  &+09 06 22 & $18.79\pm0.11$ & $15.69\pm0.08$ & $14.09\pm0.12$ & $Br\gamma$ abs & 2010-06-09, 2010-06-10\\
17&MOS07Klt15 				&19 10 17.3 & +09 06 13& $18.18\pm{0.09} $ & $15.47\pm{0.08}$  &  $14.12\pm{0.07}$  & OB & 2011-05-14, 2011-05-15\\ 
18&MOS19OBS45mid 			&19 10 17.5 & +09 06 23 & $18.81\pm{0.08}$ & $15.80\pm{0.05}$ & $14.14\pm{0.07}$ &OB & 2010-06-09, 2010-06-10\\
19&MOS11Klt145m2 			&19 10 21.9  &+09 05 03 	& $19.64\pm0.30$ & $16.01\pm0.10$  &  $14.15\pm0.11$  & $Br\gamma$ abs & 2011-05-11, 2011-05-13\\ 
20&MOS19OBS45right 			&19 10 17.5  &+09 06 24 & $19.21\pm0.14$ & $16.02\pm0.04$ & $14.34\pm0.12$ & $Br\gamma$ abs & 2010-06-09, 2010-06-10\\
21&MOS16Klt15 				&19 10 12.8  &+09 07 02& $20.41\pm{0.16} $ & $17.07\pm{0.04}$  &  $14.62\pm{0.06}$ & YSO & 2011-05-14, 2011-05-15\\
22&MOS09Klt15 				&19 10 15.9 & +09 06 05& $20.13\pm0.06$ & $16.75\pm0.08$  &  $14.70\pm0.14$  & $Br\gamma$ abs & 2011-05-14, 2011-05-15\\ 
\hline
\end{tabular}
\tablefoot{
\tablefoottext{\dag}{Stars also observed by ISAAC/VLT (PI: N. Homeier, Program ID: 073.D-0837).} 
}
\end{sidewaystable*}
\begin{sidewaystable*}
\caption{Spectroscopically observed fore- and background stars}
\begin{tabular}{ll c c c c c l l}
\hline\hline       
Source & ID & RA(J2000) & Dec (J2000) &$J$ & $H$ & $K$ & Class & Date \\ 
	&		&(h m s) & ($^{\circ}$ $'$ $''$)	&(mag) &(mag)&(mag)& & yyyy-mm-dd\\
\hline                   
23&MOS07Klt145 & 19 10 10.6 &  +09 05 06 & $17.28\pm0.01$ & $15.15\pm0.01$ & $14.13\pm0.01$  & Late Type & 2011-05-08\\
24&MOS08Klt1413 & 19 10 10.2 & +09 06 35& $18.53\pm0.02$ & $15.46\pm0.02$ & $13.91\pm0.01$  & Late Type & 2011-04-11, 2011-05-13\\
25&MOS08Klt145 & 19 10 11.2 & +09 05 33  & $17.03\pm0.01$ & $14.71\pm0.01$ &  $13.63\pm0.01$ & Late Type & 2011-05-08\\
26&MOS09Klt1415 & 19 10 16.5 & +09 07 04 &$17.10\pm0.01$ & $14.73\pm0.01$ &  $13.60\pm0.01$ & Late Type & 2011-04-11, 2011-05-13\\
27&MOS10Klt135 & 19 10 18.2 & +09 06 42 &  $18.53\pm0.03$ & $14.31\pm0.01$ &  $12.19\pm0.01$ & Late Type & 2010-05-14\\
28&MOS10Klt145 &19 10 12.4 & +09 05 26  &  $18.64\pm0.02$ & $16.01\pm0.01$ &  $14.54\pm0.01$ & Late Type & 2011-05-08\\
29&MOS10Klt145m2 & 19 10 18.1 & +09 06 52 & $19.72\pm0.05$ & $16.11\pm0.01$ & $14.34\pm0.01 $& Late Type & 2011-05-11, 2011-05-13\\
30&MOS10Klt15 & 19 10 11.1 & +09 05 28 & $19.78\pm0.06$ & $16.60\pm0.02$ & $14.78\pm0.01$  & Late Type & 2011-05-14, 2011-05-15\\
31&MOS11Klt1409 & 19 10 17.7 & +09 06 51 & $15.14\pm0.01$ & $12.02\pm0.01$ &  $10.55\pm0.01$& Late Type & 2011-04-11, 2011-05-13\\
32&MOS12Klt145 & 19 10 15.2 & +09 06 25  & $17.58\pm0.02$ & $15.37\pm0.01$ & $14.32\pm0.01$  & Late Type & 2011-05-08\\
33&MOS13Klt145 & 19 10 16.2 & +09 06 14  & $17.05\pm0.01$ & $14.86\pm0.01$ &  $13.88\pm0.01$ & Late Type & 2011-05-08\\
34&MOS13Klt145m2 & 19 10 15.8 & +09 05 38 & $17.70\pm0.02$ & $15.41\pm0.02$ & $14.43\pm0.02$ & Late Type & 2011-05-11, 2011-05-13\\
35&MOS14Klt135 & 19 10 12.2 & +09 05 37 & $18.07\pm0.02$ & $14.93\pm0.01$ &  $13.44\pm0.01$ & Late Type & 2010-05-14\\
36&MOS14Klt1408 & 19 10 18.1 & +09 06 06 & $14.17\pm0.01$ & $11.99\pm0.01$ & $11.04\pm0.01$ & Late Type & 2011-04-11, 2011-05-13\\
37&MOS14Klt145m2 & 19 10 14.8 & +09 06 34 & $18.25\pm0.02$ & $15.54\pm0.01$ & $14.23\pm0.01$ & Late Type & 2011-05-11, 2011-05-13\\
38&MOS15Klt135 & 19 10 11.6 & +09 05 27 & $16.32\pm0.01$ & $13.96\pm0.03$ & $12.93\pm0.03$ & Late Type & 2010-05-14\\
39&MOS15Klt1416 & 19 10 18.5 & +09 05 49 & $17.21\pm0.02$ & $14.92\pm0.02$ & $13.96\pm0.01$ & Late Type & 2011-04-11, 2011-05-13\\
40&MOS15Klt145m2 & 19 10 11.1 & +09 05 12 & $17.71\pm0.01$ & $15.38\pm0.01$ &  $14.27\pm0.01$& Late Type & 2011-05-11, 2011-05-13\\
41&MOS15Klt15 & 19 10 14.9 & +09 05 49 & $17.53\pm0.01$ & $15.17\pm0.01$ &  $14.13\pm0.01$& Late Type & 2011-05-14, 2011-05-15\\
42&MOS17Klt145 & 19 10 19.7 & +09 06 30 & $17.74\pm0.01$ & $15.38\pm0.01$ & $14.28\pm0.01$ & Late Type & 2011-05-08\\
43&MOS18Klt1414 & 19 10 21.3 & +09 04 58  & - & - & - &  Late Type & 2011-04-11, 2011-05-13\\
44&MOS18Klt145 & 19 10 20.4 & +09 06 39 & $18.17\pm0.01$  & $15.76\pm0.01$ & $14.54\pm0.01$ & Late Type & 2011-05-08\\

\hline
\label{Alltable}                  
\end{tabular}
\end{sidewaystable*}

\end{document}